Chiral Porphyrin Monolayers on Ferromagnetic Thin Films: Ultrafast Spectroscopy of Hybrid Interfaces

*Karol Hauza, Anna Lewandowska-Andralojc*, Ruslan Salikhov, Jürgen Lindner, Gotard Burdzinski, Marcin Kwit, Bronislaw Marciniak and Aleksandra Lindner**

K. Hauza, A. Lewandowska-Andralojc, M. Kwit, B. Marciniak
Faculty of Chemistry, Adam Mickiewicz University, Uniwersytetu Poznanskiego 8, 61-614 Poznan, Poland
E-mail: alewand@amu.edu.pl

K. Hauza, B. Marciniak
Center for Advanced Technologies, Adam Mickiewicz University, Uniwersytetu Poznanskiego 10, 61-614 Poznan, Poland

R. Salikhov, J. Lindner, A. Lindner
Helmholtz-Zentrum Dresden-Rossendorf, Institute of Ion Beam Physics and Materials Research, Bautzner Landstraße 400, 01328 Dresden, Germany
E-mail: a.lindner@hzdr.de

G. Burdzinski
Faculty of Physics and Astronomy, Adam Mickiewicz University, Uniwersytetu Poznanskiego 2, 61-614 Poznan

Funding: This research was financially supported by the National Science Centre (project no. 2020/39/I/ST5/00597) and the German Research Foundation (DFG, Deutsche Forschungsgemeinschaft, project no. 464974971).

**Abstract:** Hybrid ferromagnetic metal/organic interfaces (spinterfaces) exhibit unique properties, including spin filtering. In parallel, chiral organic molecules can themselves induce efficient spin filtering, leading to unexpectedly high spin polarizations. Here, we investigate how the proximity of gold-capped Co/Ni ferromagnetic multilayers influences the spectroscopic properties and photoinduced electron dynamics of chiral oligopeptides bearing a porphyrin chromophore. The molecules are covalently attached to the gold cap via a chiral linker, forming a self-assembled monolayer. The porphyrin macrocycles adopt an orientation parallel to the surface, resulting in the formation of J-like aggregates. Photoinduced dynamics are probed using femtosecond pump–probe transient absorption spectroscopy. Despite excitation of only a single molecular layer, a clear transient absorption signal of the porphyrin singlet excited state is observed. Adsorption on the metal surface leads to a pronounced reduction of the excited-state lifetime. However, no signatures of long-lived photoinduced charge-transfer products are detected. Furthermore, no dependence of the excited-state dynamics on either the magnetization direction of the ferromagnetic layer or the molecular chirality is observed.

## 1 Introduction

Within recent years the so called *spinterface*—the interface between ferromagnetic and molecular materials—gained a lot of attention as an ideal platform for creating new spin-related effects.[1–3] For example, it has been shown that the magnetic moments of paramagnetic molecules can be controlled via interaction with magnetically ordered substrates.[1–5] Moreover, reversible manipulation of the magnetic coupling between the adsorbed molecules and the ferromagnetic substrate can be achieved via secondary adsorption and thermal desorption of other small organic molecules onto the already existing molecular layer.[6] Furthermore, efficient spin injection (spin filtering) across the ferromagnetic metal/organic molecule interface was observed and successfully employed in prototype spintronic devices (magnetic tunnel junctions and spin valves).[7–9] In the underlying ferromagnetic film, on the other hand, the presence of the organic molecules was proven to induce magnetization hardening, softening, pinning or even modulation of the Curie temperature.[8,10–13]

These effects are attributed to the formation of the earlier mentioned *spinterface*. Once in the proximity of the ferromagnet (FM), molecules can either weakly bind to its surface via van der Waals forces (physisorption; adsorption energy less than 0.1 eV, molecule-FM surface distance greater than 3 Å, polarization of the adsorbate and surface) or strongly via covalent bond (chemisorption; adsorption energy larger than ca. 0.5 eV, the distance between the molecule and the FM less than ca. 2.5 Å, charge transfer between the molecules and the surface).[8,14] The strength of the interaction determines to what extent the electronic structure of the molecule and the metal surface will be changed. While in case of the physisorption only weak orbital overlap between the molecule and the metal occurs, which results in slight broadening and shift of the molecular orbitals energy with respect to the Fermi energy of the metal, in case of the chemisorption strong hybridization of the spin-polarized *d*- or *f*-bands of the FM and molecular orbitals occurs, giving rise to the spin-polarized (spin-split) *hybrid interface states* (HIS) responsible for the unique electronic and magnetic properties of such interfaces. In other words, initially spin-degenerated molecular orbitals split with two different energies and two different broadening width for the two spin directions (spin up ↑ and spin down ↓).[8] Interestingly, in case of a weak chemisorption, for example at the second-layer molecules which are not in a close contact with the FM surface, the interface states are located far away from the Fermi energy. In contrast, for the molecules chemisorbed in a first layer, i.e. having direct contact to the FM, the spin-split HIS are formed near the Fermi energy, driven by the hybridization of the orbitals and charge transfer interactions between molecules and the FM.[7,8] Regardless of whether it is metallic, resistive, or a dynamic spin-filtering mechanism, these are the HIS, which are responsible for the spin polarization at the spinterface.[8]

One of the possible scenarios for obtaining a well-defined spinterface is the spontaneous self-organization of molecules on the FM metal surface into well-ordered arrays with short- and/or long-range order (self-assembled monolayer, SAM) upon exposure of the surface to molecules with chemical groups that possess strong affinities for the substrate. How well these assemblies are ordered depends on the nature of the chemical interaction between the substrate and the molecules, as well as the type and strength of the intermolecular interactions among the molecules themselves.[15–17] By now, SAMs are without a doubt, among the most powerful tools for modifying the surface properties of various inorganic materials and for fabricating hybrid organic/inorganic interfaces.

Even though alkanethiols and dialkanethiols were historically used for SAM fabrication on gold, oligopeptides are also well known to form stable, densely packed films on noble metal surfaces.[18,19] Peptide-based SAM offer relative easy functionalization, possibility of anchoring on the Au surface via thiol group, disulfide group or the thiol side chain of cysteine, control over the secondary structure of the oligopeptide chain via an appropriate sequence of the amino acids as well as numerous intermolecular interactions stabilizing SAM, e.g. interchain N-H···O=C hydrogen bonds and interchain dipole interactions.[19,20] Meanwhile, it was reported for various α-helical peptides, that these interchain dipolar interactions (helix-helix interactions) play a more important role in terms of dictating the orientation of peptides in SAM than the anchoring group itself.[20,21] Due to the fact that the peptide bond is polar, once the peptide chain folds in a helical structure (i.e. adopts a well-defined secondary structure), the dipole moments of the peptide bonds sum up yielding a helix macrodipole oriented along the molecular axis, explaining the importance of the dipolar interactions among peptide chains in SAMs. An electrostatic field associated with the macrodipole promotes a directional long-range electron transfer from the C-terminal of the peptide (with an effective negative charge located there) to the N-terminal of the chain bearing an effective positive charge.[19,22] It is also responsible for the positive surface potential observed for the helical peptide layers attached to Au via the C-terminal.[23] To describe the electron transfer in peptides, two different mechanisms are typically used: *i*) electron tunneling (superexchange) where the rate of electron tunneling decays exponentially with an increase of the distance between the donor and the acceptor (single step, coherent, preferred for short distances) and *ii*) electron hopping (multi step with amide groups as hopping sites, incoherent, preferred for long distances). Which mechanism is operational for a given peptide SAM is dictated not only by the chain length of the peptide, but also by its secondary structure, dipole orientation, hydrogen bonding, oxidation potential, and the presence of the special amino acids.[24]

As by now numerously reported in the literature, chiral organic molecules themselves, including peptides, are capable of very efficient spin filtering, i.e. favoring either spin up or spin down electrons during the electron transfer/transport across the molecular frame, depending on the handedness of the chiral molecule and the direction of its electrical dipole moment once adsorbed on the surface. This effect is referred to as *chiral-induced spin selectivity* (CISS).[25–28] Moreover, once the chiral molecules are grafted on the metal's surface, they are also capable of changing the metal's magnetic properties[29,30] or even inducing "spontaneous magnetization" upon self-assembly of chiral oligopeptide on gold substrate, manifested by asymmetry in the reduction and oxidation rate constants measured by electrochemistry.[31]

The attractive idea of incorporating optically active moieties (chromophores) into the SAM on FM/Au provides in a long-term perspective the possibility of optical control and readout of electron spins and spin currents (photospintronics).[32,33] Even though very promising, surprisingly little has been done to use light-induced processes in organic chromophores as an external stimulus to study spin-dependent phenomena. The spin-polarized photoinduced electron transfer (photoinduced CISS) was studied using photoelectrochemical measurements and steady-state photoluminescence methods using bacteriorhodopsine and oligopeptide-bound cadmium selenide nanoparticles (QD) as chromophores anchored on Au-capped nickel or cobalt as ferromagnets.[32,34] For a system with cadmium selenide nanoparticles, it was shown that the excitation of the chromophore can flip the preferred spin orientation of the photocurrent, which demonstrated the unexplored potential of combining light with chirality-induced spin polarization experimentally. Moreover, spin-selectivity in perylenediimide-labelled DNA mediating charge transfer on the FM support was analyzed using fluorescence microscopy and for quantum dots anchored through a chiral oligopeptide to a gallium nitride using a Hall device under light illumination.[33] Time-resolved photoemission was used only once to probe the photoinduced electron transfer in quantum dots assembly where a donor was bound to an acceptor via a chiral linker.[35] Dynamic spin filtering at the Co/$Alq_3$ interface was also investigated by combining two-photon photoemission experiments with electronic structure theory.[7,36]

Nonetheless, the dynamics of the photoexcited chromophore covalently anchored via a chiral linker on the ferromagnetic metal has so far never been comprehensively studied with respect to the spin-selectivity of the photoinduced electron transfer, which is expected to be operative in such a system due to the combination of the *spinterface* and/or the *CISS effect*.

In this work, we investigate chiral self-assembled monolayers (SAMs) on gold-capped Co/Ni ferromagnetic multilayers (FM/Au) exhibiting strong out-of-plane magnetic anisotropy, which allows the magnetization to remain oriented perpendicular to the surface even in remanence, i.e., in the absence of an external magnetic field. The SAM consists of chiral oligopeptides bearing a porphyrin chromophore attached to the N-terminus, which provides optical sensitivity in the visible spectral range. The oligopeptide backbone adopts an α-helical conformation and is covalently anchored to the gold surface via the thiol group of a cysteine residue at the C-terminus. As a result, the helical macrodipole, oriented along the helix axis, points away from the FM/Au surface. Steady-state absorption spectroscopy is employed to characterize the structure and organization of the molecular layer, while femtosecond transient absorption spectroscopy is used to probe the excited-state dynamics of the SAM in close proximity to the FM/Au substrate.

## 2 Results and discussion

Photochemical properties of the porphyrin derivatives covalently bound to the FM/Au surface via chiral oligopeptide linker (both as L- and D-enantiomer) were investigated by steady-state and transient absorption spectroscopy. Please, see the *Experimental Section* for details concerning fabrication and structure of the magnetic heterostructures used in this study (FM/Au) as well as metallic stacks without ferromagnetic layer used for reference (abbreviated as Au). The hybrid structures under study are schematically depicted in **Figure 1A**. The 5-(4-carboxyphenyl)-10,15,20-(triphenyl)porphyrin was chosen as chromophore for this study because of its high molar absorption coefficient in the visible spectral range, stability under light irradiation and feasibility to undergo photoinduced electron transfer at the interface with metal.[37–40] The choice of 2-aminoisobutyric acid and alanine as building blocks of the oligopeptide linker was dictated by their known property to induce α-helix formation even by relatively short oligopeptide chains.[40] The α-helical structure of the chiral moiety enhances the spin filtering effect during the photoinduced electron transfer processes in chiral SAM on gold, as reported by M. Venanzi *et al*..[40] Presence of cysteine as C-terminal amino acid assured efficient chemisorption of the molecules on the gold surface through Au-S linkage leading to the formation of the self-assembled monolayer.[41], [42] Thiols are well known to form strong covalent bonds with noble metals, forming well-defined SAMs on their surfaces.[15,43] The tilt angle of the helix axis from the surface normal is affected by the choice of the solvent during SAM preparation, the nature of the amino acids in the oligopeptide sequence, the length of the oligopeptide chain as well as its secondary structure, where helix-helix interaction seems to affect the orientation of peptides in the SAM more than the Au-sulfur interaction.[21] Typically, peptides immobilized on Au via the C-terminus exhibit tilt angles of 40–66° with respect to the surface normal, resulting in a more flat-lying structure compared to analogous peptides attached to the Au surface via the N-terminus.[20,22,44] In the case of peptides attached to the Au surface via the C-terminus, the region connecting the molecules to the gold surface is highly negatively charged due to the partial negative charge at the C-terminal end of the α-helix (the negative dipole terminus) and the additional negative charge on sulfur arising from the strongly polarized Au–S bond. As a result, intermolecular electrostatic repulsion prevents the molecules from adopting close packing, leading to larger tilt angles and a more loosely packed molecular layer on the surface (**Figure 1B**).[37]

**A** **B**

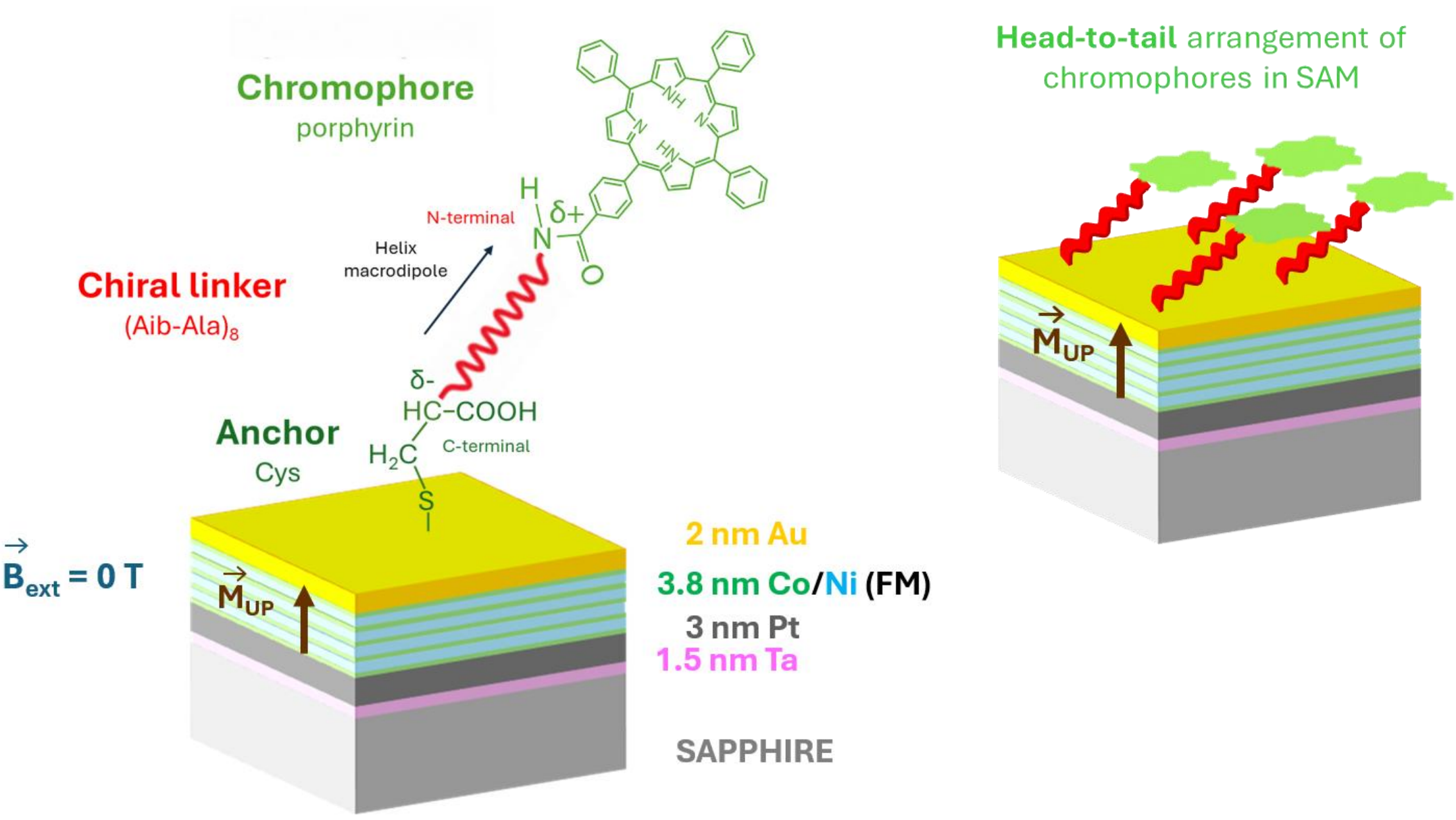

**Figure 1.** A) Schematic representation of the metallic heterostructures used in this study, together with the molecular structures employed for the formation of chiral SAMs on their surfaces. The molecular length is slightly below 5 nm, while the size of the porphyrin chromophore is approximately 1.8 nm.[39]; B) Graphical representation of an ideal SAM formed by the studied molecules on the gold-capped FM metallic layers.

## 2.1 Photophysical properties of chiral porphyrins in solution

Before the self-assembly of the organic chromophores on the ferromagnetic thin films, the spectroscopic properties of the porphyrins bearing chiral linkers were studied in solution. The enantiomeric character of the two porphyrins was confirmed by monitoring their CD spectra in ethanol (**Figure S1**). The mirror-image CD spectra show negative (positive) Cotton effects of comparable amplitudes, appearing at ca. 210 nm and 223 nm. The shape of the CD spectra is typical of the α-helical conformation of the oligopeptide chain.[39] The UV–Vis absorption spectra of L-porph and D-porph were compared with that of the reference compound TPCOOH (**Figure 2**).

The absorption spectra of L-porph and D-porph in ethanol exhibit distinct bands at 415 nm (Soret band, $S_0$–$S_2$ transition) and at 512, 546, 589, and 645 nm (Q bands, $S_0$–$S_1$ transitions).

The positions of the corresponding bands for TPCOOH are almost identical, with only a 1 nm blue shift observed for the Soret band. The presence of the oligopeptide slightly reduces the molar absorption coefficient of the Soret band (**Table 1**). At the same time, it does not alter the spectral width of any band, as shown in **Figure S2**, where the normalized spectra of all three porphyrins are overlaid.

The fact that the spectra of the L/D porphyrins closely resemble that of TPCOOH supports the conclusion that there are no specific interactions between the attached oligopeptide chains and the porphyrin macrocycles that would affect light absorption in the visible range. The similarity of the absorption spectra of all three porphyrins clearly indicates that the introduction of the chiral linker does not affect the ground-state electronic structure of the porphyrin chromophore.

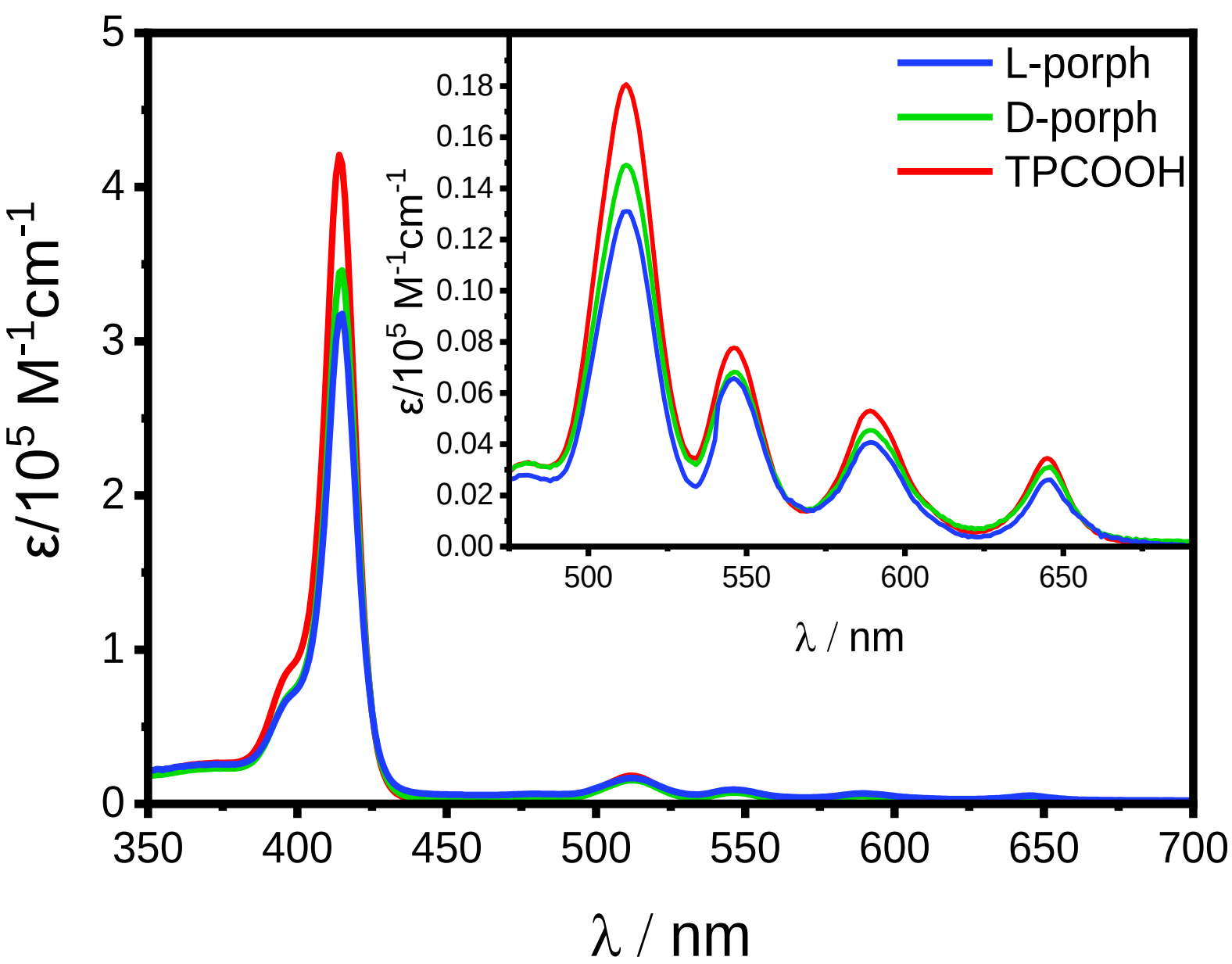

**Figure 2.** Absorption spectra of the L-porph (blue line), D-porph (green line) and TPCOOH (red line) in EtOH; (inset: Q-band region of the same spectra).

**Table 1.** UV–Vis absorption properties of the three porphyrins in solution. The extinction coefficient values are in $10^3$ $M^{-1}$ $cm^{-1}$ units. The wavelengths of Soret and Q-bands are presented in parentheses (in nm).

| Compound | Soret band $\varepsilon(\lambda_{max})$ | Q bands $\varepsilon(\lambda_{max})$ | | | |
|---|---|---|---|---|---|
| L-porph | 318(415) | 16.6(512) | 9.4(546) | 6.7(589) | 5.3(645) |
| D-porph | 347(415) | 16.7(512) | 8.5(546) | 6.2(589) | 4.8 (645) |
| TPCOOH | 421(414) | 18.1(512) | 8.0(546) | 5.5(589) | 3.5(645) |

It is well known that porphyrins readily form aggregates, in a manner strongly dependent on their concentration and the solvent used.[45–48] Therefore, prior to immersing the FM/Au and Au samples in solutions of the studied porphyrins, it was important to determine whether, at the concentrations typically used for SAM formation, our porphyrin derivatives remain in their monomeric form. In ethanol, Beer–Lambert plots of absorbance as a function of L-porph and D-porph concentration were linear up to 90 μM, indicating that the chiral porphyrins remain in their monomeric form even at relatively high concentrations (**Figure S3**). The bandwidth at half maximum ($W_{1/2}$) of the Soret band was 14 nm across the entire concentration range studied. Based on these observations, it can be concluded that the same species, namely the monomer, is present throughout the examined concentration range.

Subsequently, the excited-state properties of L-porph and D-porph were evaluated by means of steady-state and time-resolved emission techniques. The fluorescence spectra of both enantiomeric porphyrins are shown in **Figure 3A**, with selected photophysical properties summarized in **Table 2**.

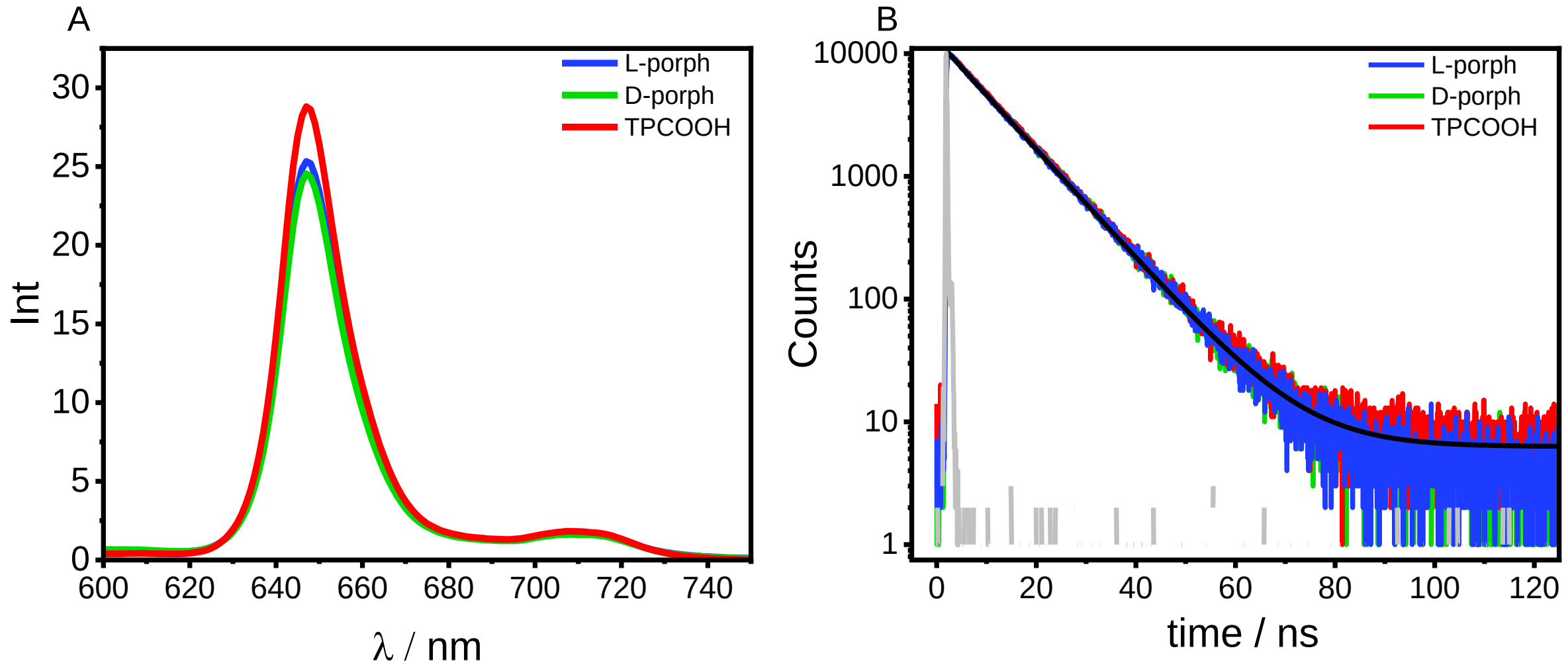

**Figure 3.** A) $S_1$-$S_0$ emission spectra of L-porph (blue line), D-porph (green line) and TPCOOH (red line) in EtOH measured at an excitation wavelength of 414 nm, B) decay of L-porph, D-porph and TPCOOH fluorescence at 647 nm recorded in EtOH with $\lambda_{ex}$ = 405 nm; grey curve – instrument response function (prompt); black curve – the monoexponential decay fit.

The emission spectra of chiral porphyrins and TPCOOH are characterized by the most intense band at λ = 647 nm and a less intense band at λ = 708 nm. The absence of a shift in the emission maximum of the TPCOOH porphyrin compared to the chiral porphyrins further confirms that the attachment of the peptide chain induces negligible changes in the electronic structure of the porphyrin core. The emission spectra of chiral porphyrins are nearly identical, which aligns with expectations for enantiomeric compounds. To verify the absence of impurities in the studied compounds, emission spectra of the individual porphyrins were recorded at different excitation wavelengths. The perfect overlap between the fluorescence excitation and absorption spectra indicates that the studied compounds do not contain impurities that fluoresce within the examined spectral range (**Figure S4**). The fluorescence quantum yield measurements were conducted using a relative method, based on a reference compound with a known fluorescence quantum yield. Determined fluorescence quantum yields are quite similar for L-porph, D-porph and TPCOOH (Table 2). The fluorescence lifetimes of all porphyrins at $\lambda_{em}$ = 647 nm were determined using the TCSPC technique (**Figure 3B**). As shown in Table 2, chiral porphyrins and TPCOOH exhibit identical fluorescence lifetimes. This indicates that the attachment of the peptide chain does not affect the lifetime of the porphyrins' singlet excited state.

**Table 2.** Comparison of the emission properties of L-porph, D-porph and TPCOOH in EtOH.

| **Compound** | $\lambda_{em}$ (nm) | $\Phi_{Fl}$ | $\tau_S$ (ns) | $k_{Fl}$ ($10^7$ $s^{-1}$) | $\Delta\lambda_{Stokes}$ ($cm^{-1}$) |
|---|---|---|---|---|---|
| L-porph | 647 | 0.23 | 9.8 | 2.3 | 47.9 |
| D-porph | 647 | 0.21 | 9.9 | 2.2 | 47.9 |
| TPCOOH | 647 | 0.25 | 9.9 | 2.5 | 47.9 |

$\lambda_{em}$ – maximum emission wavelength; $\Phi_{Fl}$ – fluorescence quantum yield; $\tau_S$ - lifetime of the singlet excited state; $k_{Fl}$ - fluorescence rate constant; $\Delta\lambda_{Stokes}$ – Stokes shift

To fully characterize the excited-state dynamics of the porphyrins in solution, ultrafast pump–probe spectroscopy experiments were performed in ethanol using a 390 nm laser pulse as the pump. This wavelength was chosen to ensure efficient excitation of the porphyrin chromophores while minimizing overlap between the pump pulse and the ground-state bleach in the Soret band region.

Immediately after excitation of L-porph or D-porph, the transient absorption spectra (**Figure 4A** and **Figure 4C**) revealed a bleach centered around 415 nm, arising from ground-state depletion as the porphyrin molecules were excited by the laser pulse. In the singlet excited state, both L-porph and D-porph exhibited a broad and intense $S_1 \rightarrow S_n$ absorption band with a maximum at 435 nm, together with a Q-band bleach that coincided with the Q-band positions observed in the steady-state UV–Vis absorption spectra.

Excitation at 390 nm corresponds to a transition from the $S_0$ ground state to a vibrationally excited level of the $S_2$ state of the porphyrins (Soret band). Analysis of the transient absorption kinetics at 435 nm at short delay times revealed a fast decay component with time constants of 8.8 ps and 9.8 ps for L-porph and D-porph, respectively (**Figure 4B** and **Figure 4D**). The time constants obtained from the recovery of the bleach at early delay times (*measured at 420 nm,* *instead of 415 nm, to exclude scattered pump light*) —11.2 ps and 10.6 ps for L-porph and D-porph, respectively—are in good agreement with the decay constants derived from the transient absorption at 435 nm. The observed decay can be attributed to $S_2 \rightarrow S_1$ internal conversion. Accordingly, this time constant can be interpreted as the lifetime of the $S_2$ excited state, in accordance with literature data available for other porphyrin derivatives.[49,50]

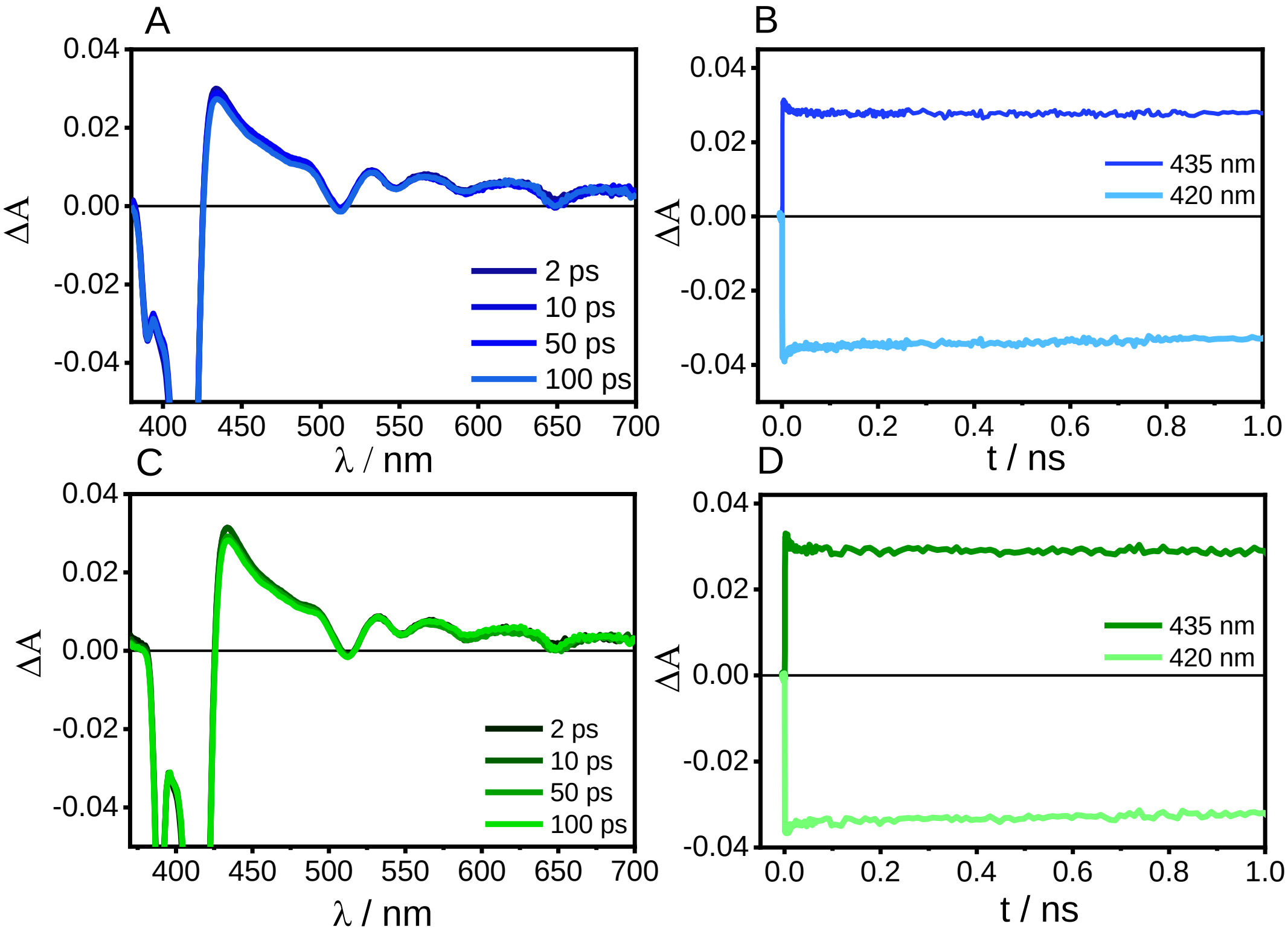

**Figure 4.** Transient absorption spectra acquired at various time delays in solution for A) L-porph and C) D-porph after 390 nm laser excitation; B) ΔA time profiles at 435 nm (dark blue curve) and 420 nm (light blue curve) registered for the L-porph in EtOH; D) ΔA time profiles at 435 nm (dark green curve) and 420 nm (light green curve) registered for D-porph in EtOH. Data probed at 390 nm are affected by scattered excitation pulse.

The $S_1$ singlet excited state lifetime of chiral porphyrins could not be determined from the femtosecond transient absorption spectroscopy due to the experiment's short time window of 3 ns, consistent with the TCSPC findings showing that this lifetime for all the studied porphyrins is ca. 10 ns, which agrees well with the literature data.[51] Notably, throughout the experiment's time window, the decay of the singlet excited state signal was not accompanied by any distinct spectral evolution.

## 2.2 Photophysical properties of chiral porphyrins in SAMs

### *2.2.1 UV-Vis Characterization*

Prior to the adsorption of chiral porphyrins, the transmission of light through the metallic heterostructures used in this study was measured (see **Figure 5A**). The measured extinction (extinction = $\log_{10}(1/T)$ corresponds to the sum of absorption and scattering losses of the thin metallic films, with absorption dominating over scattering. Elastic (Rayleigh) scattering is negligible because the studied films are continuous and exhibit low surface roughness, typically <1 nm.

In the UV–Vis-NIR region (≈ 300–800 nm), the optical response of thin metal films has two main contributions: one arising from free electrons (Drude response; intraband transitions) and another one from interband transitions. The former produces a broad, featureless extinction background that becomes dominant in the red and near-infrared regions, whereas the latter gives rise to additional features in the extinction spectra. The optical response of a metal is determined by its electronic band structure. As a result, different metals exhibit characteristic spectral behavior even for very thin films.[52–55]

In the case of the metal stacks without a ferromagnetic layer, the observed extinction is dominated by the 5d → 6sp interband transitions in gold, which begin around 500–540 nm (ca. 2.3–2.5 eV)[56,57] This manifests itself as an increase in extinction below 500 nm. Additional contributions arise from interband transitions in Pt, which lead to an almost monotonic increase in extinction toward the UV across the entire visible range.[58]

For the metal stacks containing the ferromagnetic metals Co and Ni, the extinction values increase significantly across the entire analyzed spectral range, and the overall spectral shape changes slightly. This behavior can be attributed to the 3d → 4sp interband transitions in these metals.[52]

The transmission of light through the metallic heterostructures was measured again after chemisorption of the chiral porphyrins forming SAMs on the gold surface. The Soret band of L-porph and D-porph in the SAMs is red-shifted by 6 nm relative to the absorption spectra of the corresponding molecules in ethanol (**Figure 5B**).

The increase in extinction at the Soret band maximum upon SAM formation arises solely from absorption of light by the molecules covalently linked to the gold capping layer and therefore directly reflects the absorbance of the SAMs. Because the absorbances of both chiral SAMs are similar regardless of the presence of the ferromagnetic layer (**Table 3**), the affinity of the two enantiomers for the gold surface must be comparable in all cases. The observed absorbance values are also consistent with those reported for other porphyrins arranged in SAMs on gold surfaces.[59] Moreover, their relatively large magnitude provides additional information about the orientation of the porphyrin transition dipole moments. The strength of optical absorption depends on the projection of the molecular transition dipole moment onto the electric field of the incident light, and therefore varies with molecular orientation within the monolayer.[60–62] In the studied SAMs, the observed absorbance is consistent with a nearly flat orientation of the porphyrin macrocycles relative to the surface[60,63] which also agrees well with the J-type interactions inferred from the red shift of the Soret band, as discussed further on.[64]

Moreover, the ground-state absorption spectra of the L/D-porphyrin SAMs on Au (without a ferromagnetic layer) are identical to those on FM/Au not only in terms of the absorbance at the Soret band (Table 3) but also with respect to the band width and peak position (Figure 5B, Table 3). This indicates that the presence of the ferromagnetic layer beneath the thin gold film neither affects the affinity of the porphyrins for the gold surface, as discussed above, nor modifies their ground-state electronic structure. Within our sensitivity limits, the out-of-plane magnetization direction of the ferromagnetic layer (UP vs DOWN) does not affect the ground state absorption spectra of the L/D -porphyrin SAMs either (**Figure S5**).

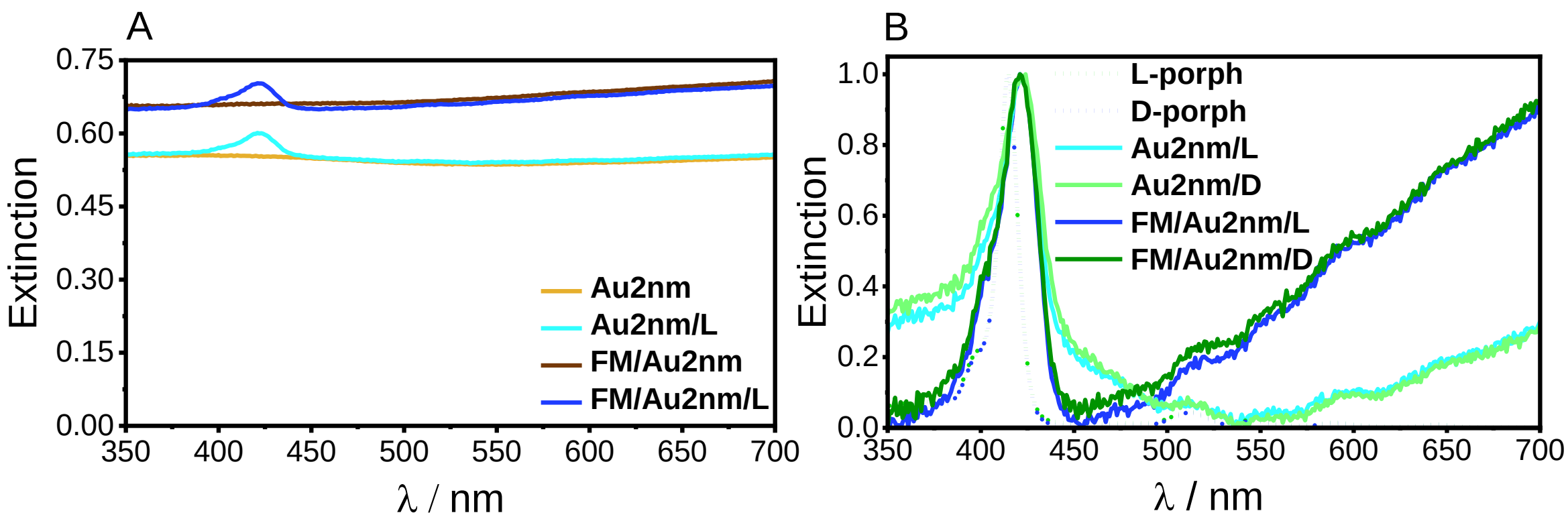

**Figure 5.** A) Extinction spectra of the metallic heterostructures measured before and after adsorption of the molecules. B) Comparison of the normalized UV–Vis absorption spectra of the porphyrins in ethanol with the normalized extinction spectra of the corresponding SAMs on FM/Au and Au substrates.

The bandwidth at half maximum ($W_{1/2}$) of the Soret band is 32 nm for the L-porph SAM on FM/Au and 34 nm for the D-porph SAM, compared with 14 nm for the same porphyrins in ethanol. For the SAMs formed on FM/Au, the corresponding $W_{1/2}$ for L-porph and for D-porph are 30 nm. Similar broadening of the Soret band, accompanied by a shift toward longer wavelengths, has previously been reported for SAMs of other porphyrin derivatives on gold substrates[59,65–68] and for porphyrin-containing Langmuir–Blodgett films on glass.[69–71] These effects are attributed to π–π interactions between closely packed porphyrin macrocycles. In particular, it is well established that head-to-tail aggregation of chromophores (formation of J-aggregates) leads to a red shift of their absorption bands (see **Scheme S1**).[72] Following this interpretation, the spectral changes observed upon SAM formation are attributed to head-to-tail π–π interactions between neighboring porphyrin macrocycles within the monolayer (see Figure 1B), rather than to the formation of large, extended J-aggregates typically observed in solution.[73]

Such interactions are consistent with a molecular orientation in which the porphyrin macrocycles lie approximately parallel to the gold surface, enabling head-to-tail coupling between adjacent chromophores.[64] [74] In contrast, a perpendicular orientation of the macrocycles relative to the surface would favor a face-to-face configuration and thus H-type aggregation.[60,68]

**Table 3.** UV–Vis spectroscopic parameters and molecular surface concentrations in SAMs.

| Sample | $\lambda_{max}$ (nm) | Abs. | $W_{1/2}$ | Ƭ (molecules cm$^{-2}$) |
|---|---|---|---|---|
| Au-L-porph | 421 | 0.045 | 30 | $5.6 \times 10^{13}$ |
| Au-D-porph | 421 | 0.043 | 30 | $5.4 \times 10^{13}$ |
| FM/Au-L-porph | 421 | 0.042 | 32 | $5.3 \times 10^{13}$ |
| FM/Au-D-porph | 421 | 0.038 | 34 | $4.7 \times 10^{13}$ |

$W_{1/2}$ - bandwidth at half maximum; Ƭ - surface concentration of the molecules

These two different chromophore orientations are associated with markedly different surface packing densities of the molecules. Van Galen and Majda, using a simple square lattice model, predicted that for 10,15,20-tetra(4-pyridyl)porphyrin (CoTPyP) monolayers the surface concentration would be approximately $4.2 \times 10^{13}$ molecules cm$^{-2}$ for the parallel arrangement and $20 \times 10^{13}$ molecules cm$^{-2}$ for the perpendicular orientation.[75] Similarly, Shimazu *et al.* studied two porphyrins - Zn(II) 5-[p-(6-mercaptohexoxy)phenyl]-10,15,20-triphenylporphyrin (ZnP1S) and Zn(II) 5,10,15,20-tetrakis-[p-(6-mercaptohexoxy)phenyl]porphyrin (ZnP4S) - which are believed to adsorb onto gold surfaces in predominantly perpendicular (face-to-face, ZnP1S) and parallel (head-to-tail, ZnP4S) orientations, respectively. Based on UV–Vis measurements, surface densities of $17 \times 10^{13}$ molecules cm$^{-2}$ for ZnP1S (perpendicular) and $6.6 \times 10^{13}$ molecules cm$^{-2}$ for ZnP4S (parallel) were reported.[64]

These examples demonstrate that porphyrins arranged in a face-to-face (perpendicular) orientation generally exhibit significantly higher surface packing densities than those aligned head-to-tail (parallel) with respect to the surface.[76] Therefore, the experimentally determined surface concentrations can provide useful insight into the relative orientation of the chromophores within the SAM.

Using the model proposed by Li *et al.*, which predicts that the molar absorption coefficient of a chromophore in a monolayer is 3/2 times that of the same chromophore in solution,[77] the molecular surface concentrations were estimated by comparing the absorption spectra of the chiral porphyrins in solution with those of the corresponding SAMs, having the absorbances matched at the Soret band. The obtained surface coverages were similar for both enantiomers and were likewise not significantly affected by the composition of the metallic heterostructures (see Table 3). Such relatively low coverages are indicative of a predominantly parallel orientation of the porphyrin chromophores with respect to the surface[65,66,68] and thus support our earlier conclusion of head-to-tail interactions between neighboring macrocycles, inferred from the red shift of the Soret band. They are also in agreement with the expected looser packing of the oligopeptide chains and larger tilt angles from the surface normal expected for α-helical peptides bound to the gold via their C-terminus, as explained at the beginning of this section.

The chemisorbed molecules are likely organized into domain-like regions that exhibit slight variations in tilt angle, tilt direction, and even chromophore orientation, rather than forming a perfectly uniform monolayer.

*2.2.2. Femtosecond transient absorption spectroscopy*

Ultrafast pump–probe absorption spectroscopy is well-suited to probe excited-state dynamics and to detect possible charge-transfer products. As mentioned in the Introduction, the proximity of a ferromagnet, as well as the presence of chiral elements in SAMs, can lead to spin polarization. Our metallic heterostructures with chiral SAMs were designed so that reversing the direction of the ferromagnet magnetization (UP *versus* DOWN) could differently affect the photoexcited electron dynamics depending on the chirality of the SAM, provided that electron transfer is involved in the relaxation of the photoexcited states.

Chemisorption of the chiral L-porph and D-porph molecules occurs via hybridization of the sulfur p orbitals with the d orbitals of gold, while the porphyrin HOMO and LUMO remain largely unaffected, as shown by the steady-state absorption measurements. Although the magnetic proximity effect can, in principle, induce spin polarization in noble metals, the resulting magnetic moments are confined to only a few atomic layers at the interface with a ferromagnet.[78] Consequently, for a gold capping layer of finite thickness, any exchange-induced band splitting at the Au/SAM interface is expected to be largely diminished.[79] Nevertheless, a small residual spin polarization may still persist.

The chiral oligopeptide linkers themselves, however, are expected to preferentially transmit electrons with a given spin orientation depending on their chirality (CISS effect).[28] Consequently, once photoinduced electron transfer occurs, spin polarization may build up in the system.

The first requirement for efficient photoinduced electron transfer (PET) between an organic chromophore and a thin metallic film (in either direction) is favorable energetics.[80] Appropriate energy level alignment is necessary to ensure sufficient driving force for electron transfer to occur. The Fermi level of gold lies at approximately −5.1 eV relative to the vacuum level,[81] corresponding to about + 0.66 V vs the Standard Hydrogen Electrode (SHE). This value is derived from the commonly accepted work function of gold and the standard conversion between electrochemical and vacuum energy scales.[82,83] It should be noted that the exact position of the Fermi level may vary depending on the surface structure, adsorbed molecular layers, and interfacial dipoles.[81] Using the experimentally determined oxidation and reduction potentials of the porphyrin chromophore bearing α-helical substituent with analogous amino acid sequence as this used in our studies and arranged in SAM on gold surface ($E_{ox}(Porph^{\bullet +}/Porph)$ = 0.85 V vs SHE; $E_{red}(Porph/Porph^{\bullet -})$ = −1.05 V vs SHE),[39] together with $E_{00}$ determined from the intersection point of the normalized absorption and emission spectra. ($E_{00}$ = 1.97 eV), we could estimate the driving force for electron transfer from the photoexcited porphyrin to gold based on the Rehm-Weller equation:[61]

$$\Delta G_{electron\,transfer} = E_{ox}(Porph^{\bullet +}/Porph) - E_F(Au) - E_{00}$$

Even though the obtained value of −1.78 eV can only be considered as a rough estimation, its large negative value indicates that the electron injection from the photoexcited porphyrin chromophores into Au, at least from the point of view of thermodynamics, should be highly favorable, i.e. the excited electron should have sufficient excess energy above the metal Fermi level to inject into the metal continuum.

In contrast to electron transfer from the excited chromophore to the metal, the driving force for hot-electron injection from gold into the porphyrin cannot be described by a single thermodynamic parameter, as hot carriers in the metal exhibit a non-equilibrium energy distribution. Nevertheless, considering the photon energies used in these studies (2.95 eV and 3.54 eV, which correspond to 420 nm and 350 nm) and typical reported energies of LUMO of tetraphenylporphyrin and its derivatives (−3.3 to −3.6 eV),[84] such injection is energetically feasible, although expected to be rather inefficient due to strong competition from ultrafast hot-electron relaxation in the metal and limited electronic coupling across the chiral linker.[85]

Even in the presence of favorable energetics, the overall efficiency of PET is further controlled by electronic coupling across the interface dependent on porphyrin orientation and the distance imposed by the oligopeptide linker, which may significantly limit the electron transfer rate, as well as all the ultrafast competing relaxation processes in both the molecule/SAM and the metal.[80]

In this context, it is essential to directly probe the excited-state dynamics at such hybrid interfaces. To the best of our knowledge, this is the first time that femtosecond transient absorption (TA) spectroscopy has been applied to investigate self-assembled monolayers of chiral porphyrins on gold-capped thin ferromagnetic films (FM/Au/SAM). As a reference system, SAMs of these porphyrins on gold without an underlying ferromagnetic layer were also studied (Au/SAM).

To probe the influence of the FM magnetization direction on the porphyrin excited-state dynamics in SAMs, the TA spectra were recorded twice for each sample, switching the magnetization direction of the FM layer (UP *versus* DOWN) between the measurements.

As shown in **Figure 6A** and **Figure 6C**, excitation of the reference samples with a 2 nm gold capping layer, both without and with the FM layer, using a 420 nm pump pulse results in a broad and largely featureless positive transient absorption signal across the probed spectral range of 440–650 nm. This signal decays at longer pump–probe delay times. A closer inspection of this spectral region reveals a weak absorption band centered around 500–520 nm. The intensity of this band increases with increasing gold thickness, and a gradual blue shift with increasing delay time becomes apparent (see **Figure S6**).

The observed transient absorption originates from interband 5d → 6sp transitions in gold and the generation of non-thermalized hot electrons. Optical excitation leads to rapid heating of the electron gas, which transiently modifies the dielectric function of the metal and distorts the interband transition edge. As the electron gas cools, this distortion relaxes, giving rise to the characteristic blue shift of the transient absorption band.[86–89]

Thermalization of hot electrons via electron–electron scattering occurs on a sub-picosecond timescale, followed by cooling of the thermalized hot electrons on a 1–3 ps timescale through electron–phonon coupling, i.e., energy transfer from hot electrons to the gold lattice. Subsequently, over ~ 200–300 ps, the system returns to thermal equilibrium through heat dissipation into the substrate via phonon–phonon coupling, as illustrated in Scheme 1.[89],96

Decay kinetics for the reference samples, recorded at 445 nm and 500 nm following 420 nm excitation, are shown in **Figure S7**. Despite the relatively low signal-to-noise ratio, a clear formation of a positive signal at 500 nm is observed for 2 nm Au within the pump pulse, which is attributed to hot electrons. This signal decays within the first ~ 2 ps, consistent with energy dissipation into the Au lattice, and is followed by a broader signal that reaches a maximum at 10–20 ps and decays to zero on a timescale of ~ 500 ps. For the FM/Au2nm sample, no distinct hot-electron signal is observed immediately after the pump pulse at 500 nm. Nevertheless, a positive signal builds up, reaching a maximum at approximately 10–20 ps and decaying within ~ 500 ps. This behavior is attributed to heating of the metallic heterostructure. As illustrated in **Figure 6B** and **Figure 6D** for L-porph (see **Figure S8** for D-porph), the formation of a SAM on the gold surface strongly affects the transient absorption response of the studied systems, both in the absence and in the presence of the FM layer. In contrast to the solution measurements, the SAMs were excited with a 420 nm pump pulse (instead of 390 nm) to ensure the highest possible direct excitation of the porphyrin chromophores at the Soret band.

However, analysis of the extinction spectra (see Supporting Information and **Table S1**) shows that approximately 95–96% of the absorbed pump energy (for 420 nm excitation) is dissipated within the metallic stack, while only 4–5% is absorbed by the molecular layer. The excitation is therefore dominated by the metal, and direct excitation of the SAM constitutes only a minor contribution. Nevertheless, it can still give rise to a detectable ground-state bleach at the porphyrin Soret band as well as the Q bands, along with positive transient absorption signals, owing to the large molar absorption coefficients of the porphyrin chromophores[91,92] as evidenced by the transient absorption spectra of Au/SAM and FM/Au/SAM (**Figure 6** and **Figure S8**). At the same time, indirect excitation of the SAM, mediated by interfacial energy transfer from the metal stack (see Scheme 1) is likely operative as well.

TA spectra following 420 nm excitation of FM/Au/SAM ($S_0 \rightarrow S_2$ transition in chromophores) closely resemble those of the corresponding porphyrins in solution (cf. Figure 4 and Figure 6D). A ground-state bleaches centered around Soret band and Q bands, indicating depletion of the porphyrin ground state, are overlapped with a positive transient signal having a maximum at approximately 445 nm, being only ~10 nm red-shifted compared to the porphyrin TA spectra in solution. Importantly, neither the chirality of the SAM (L vs D; see Figure 6 and Figure S8 as well as **Figure 8** for the transient absorption kinetics at 445 nm) nor the magnetization direction of the ferromagnetic layer (compare Figure 6D, Figure S8D, and **Figure S9**) significantly affect the TA spectra. It should be noted, that the ground-state bleach at ~ 420 nm partially overlaps with a negative signal that is independent of the pump–probe delay. This delay-independent contribution is attributed to residual pump scattering reaching the detector rather than to a true ground-state bleach, as it does not follow the temporal evolution of the excited-state population. The positive transient absorption signal is assigned to the porphyrin $S_1$ excited state ($S_1 \rightarrow S_n$ transitions). It is formed rapidly via internal conversion from the initially populated $S_2$ state upon direct photoexcitation of porphyrins.[93,94] In parallel, the porphyrin chromophores can be indirectly excited via energy transfer from the metal. No evidence for the formation of charge-transfer products was observed.

In contrast to porphyrins in solution, where the excited state persists on the nanosecond timescale[95] the excited state of porphyrins adsorbed on the metallic heterostructure decays within the first ca. 50 ps after the pump pulse (Figure 7B). The processes responsible for this rapid relaxation will be discussed later.

Excitation of Au/SAMs with a 420 nm pump yields different transient absorption spectra than those of FM/Au/SAMs. The overall spectral shape is more featureless, however, at short time delays, bleaches corresponding to the porphyrin Q-bands suggest a contribution from the porphyrin $S_1$ state. At long time delays, the TA signal seems to be dominated by the photoexcited electrons in gold with a maximum around 500 nm.

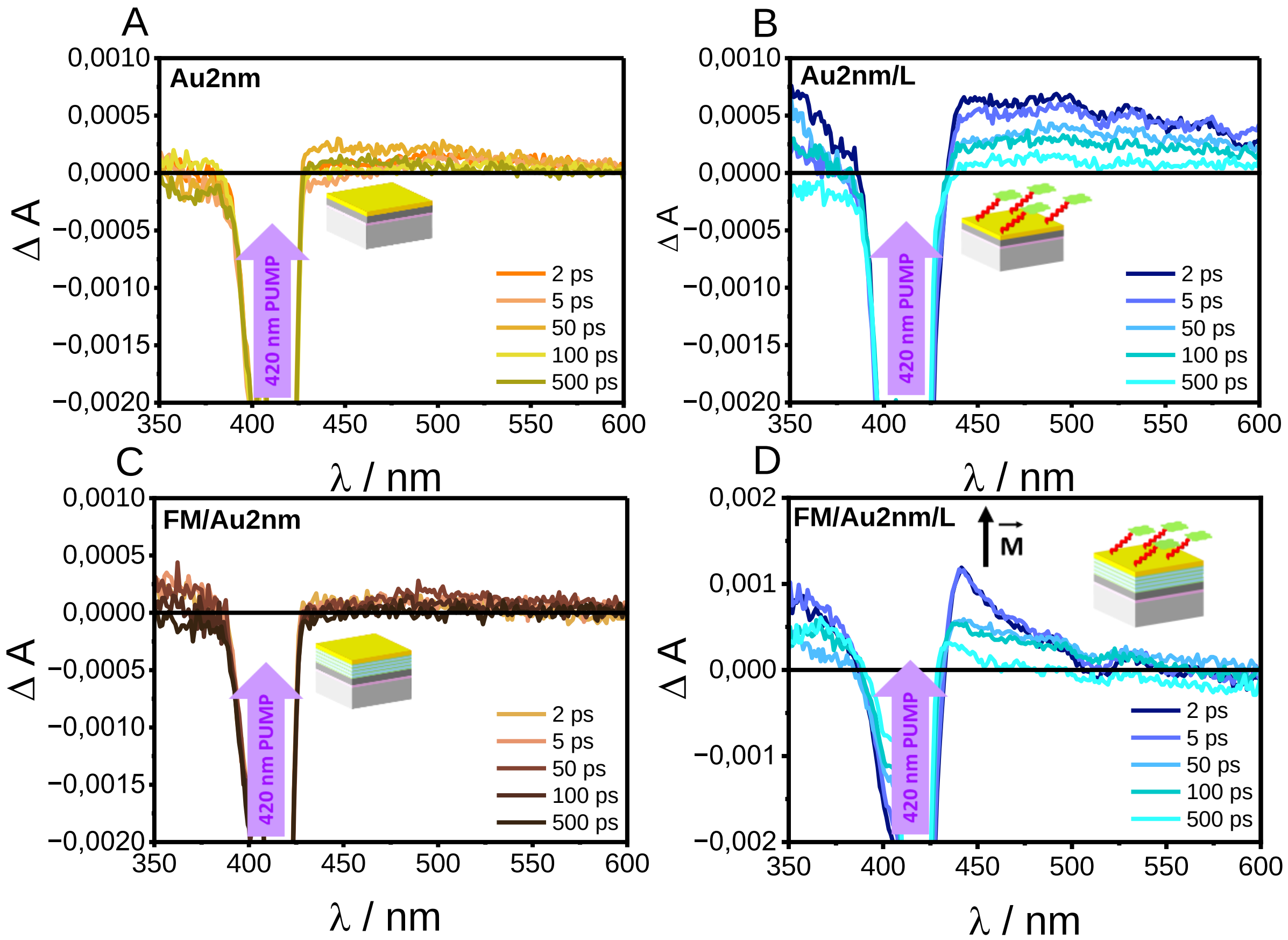

**Figure 6** Transient absorption spectra registered at various time delays for two reference samples without SAM: A) Au and C) FM/Au as well as corresponding samples with adsorbed molecules: B) Au-L-porph and D) FM/Au-L-porph with magnetization "up" following the 420 nm excitation.

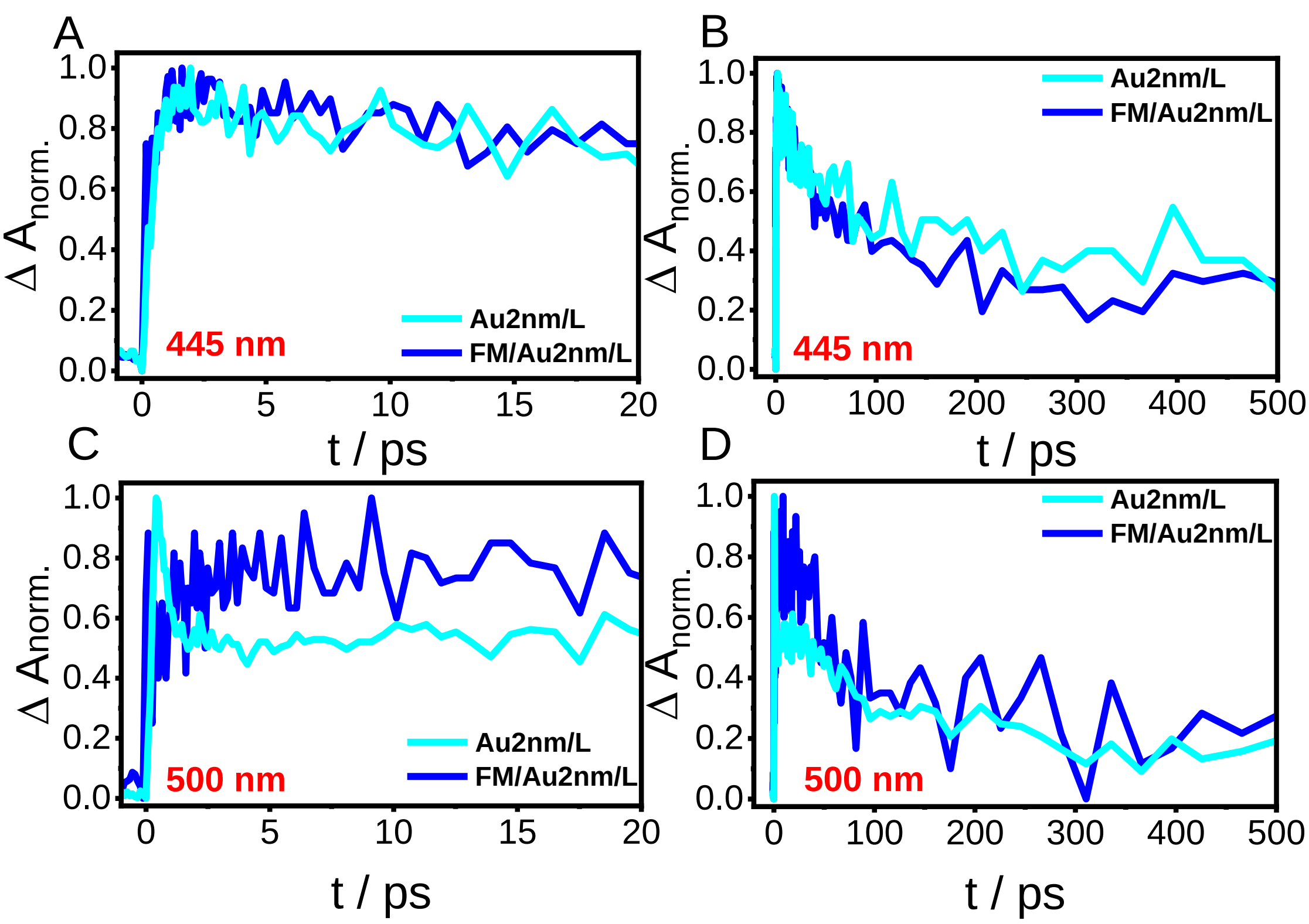

**Figure 7.** Transient absorption kinetics at 445 nm and 500 nm on short- and long-time scale for metallic heterostructures with L-porph anchored to the gold surface following the 420 nm laser excitation. Samples with a ferromagnetic layer were magnetized "up".

As shown in **Figure 7A** and **Figure 7B**, both the initial rise of the TA signal at 445 nm and its subsequent decay are essentially identical for Au/SAM and FM/Au/SAM. The rise is completed within approximately 2 ps, while the decay can be well described by a biexponential function, yielding a fast component of ~ 20–40 ps and a slower component of ~ 200–260 ps (for precise values see **Table S2**).

A different behavior is observed at 500 nm. As illustrated in **Figure 7C**, for the Au/SAM samples the TA signal is already generated within the duration of the pump pulse. This initial component decays within the first ~2 ps (Figure 7C) and simultaneously the growth of a second transient species is observed, which reaches its maximum at around 10–20 ps (**Figure 7D**). In contrast, for FM/Au/SAM the component formed within the pump pulse duration is significantly less pronounced. Despite these differences in the early-time dynamics, the decay of the longer-lived TA signal for both Au/SAM and FM/Au/SAM occurs on comparable timescales, resembling the long-time decay of the signal at 445 nm.

The above observations also hold for samples bearing SAMs of opposite chirality (see F**igure S10**). A compilation of the decay constants obtained from fitting the transient absorption kinetics at 445 nm is presented in Table S2 and shows that neither SAM chirality nor the ferromagnet magnetization direction systematically influences the observed relaxation processes. Owing to the relatively low signal-to-noise ratio, the extracted decay constants should be regarded as equivalent within experimental uncertainty, both for the fast and slow components. Hence, the presence of a fast component (~20–40 ps) and a slower component (~200–260 ps) can be reliably identified.

To elucidate the origin of the differences observed in the transient absorption spectra of Au/SAM and FM/Au/SAM (Figure 6), as well as the nature of the transient formed within the pump pulse duration and detected at 500 nm (Figure 7C), the pump wavelength was changed to 350 nm. TA spectra registered for the reference samples without molecules are depicted in **Figure S11**. As shown in Fig. 8, directing even more excitation into the metallic heterostructure (see Table S1; lower absorbance of SAM at 350 nm then at 420 nm) clearly highlights the differences between samples without and with the ferromagnetic layer. In the absence of the FM (**Figure 8A** and **Figure 8B**), the TA spectra at short time delays exhibit a pronounced, broad maximum around 500 - 520 nm, accompanied by a bleach in the Soret band region with a shoulder at approximately 460 nm.

The TA maximum is attributed to hot electrons in the gold layer. It undergoes a characteristic blue shift with increasing delay time, reflecting the cooling of the hot electron gas via electron–phonon relaxation in Au and the gradual recovery of the equilibrium interband dielectric function of gold.[86–89]

The presence of the porphyrin ground-state bleach indicates that a fraction of the chromophores is in the excited state, either as a result of direct optical excitation or via interfacial energy transfer from the metal, as illustrated in Scheme 1.

Interestingly, when the FM layer is present (**Figure 8C** and **Figure 8D**), the generation of hot electrons is strongly suppressed. Instead, the overall TA spectral shape more closely resembles that of the porphyrins, exhibiting a ground-state bleach around 420 nm and a TA maximum at approximately 445 nm. The reduced hot-electron signal observed for FM/Au/SAM compared to Au/SAM is attributed to an additional ultrafast energy dissipation pathway enabled by the presence of the ferromagnetic Co layer [96]. In the Au/(Co/Ni) heterostructure, laser-excited hot ***sp***-electrons in the Au layer undergo enhanced scattering with localized Co ***d***-electrons, opening an additional relaxation channel that strengthens electron–phonon coupling and accelerates energy transfer from the electronic system to the lattice[96]. As a result, the electron temperature in the Au layer decays more rapidly, leading to a suppression of the hot-electron population and, consequently, a weaker hot-electron transient absorption signal compared to the Au/SAM system. Taking this into account, it is reasonable to conclude that the porphyrin chromophores are excited predominantly directly by the pump pulse, rather than indirectly via hot-electron-mediated interfacial energy transfer.

The TA kinetics obtained following 350 nm excitation presented in **Figure 9** (see the respective data in Figure S12 for the D-enantiomer) resemble, to some extent, those shown in Figure 7, obtained following 420 nm excitation. At 445 nm, the initial rise of the positive signal within the first few picoseconds after the pump pulse is again followed by a biexponential decay. However, this rise is superimposed on a bleach at the band edge of the interband transitions in gold, which forms within the pump pulse and is more efficiently excited by 350 nm irradiation than by 420 nm irradiation. At the same time, the slower component of the TA decay for Au/SAM (~ 340–360 ps) differs from that observed for FM/Au/SAM (~ 200–250 ps), demonstrating that Co-layer-mediated electron–phonon equilibration leads to an even faster phonon relaxation in the FM-based system.

At 500 nm, for the Au/SAM the relative amplitudes of the signal generated within the pump pulse and the one reaching its maximum at 10–20 ps change compared to excitation with a 420 nm pump (cf. Figure 7C and Figure 9C). The more slowly rising component increases in intensity relative to the prompt signal, which can be attributed to greater heat deposition in the metallic heterostructure when using the 350 nm pump. The long-time decay of the signal at 500 nm appears to follow similar kinetics regardless of the presence of the FM layer.

.

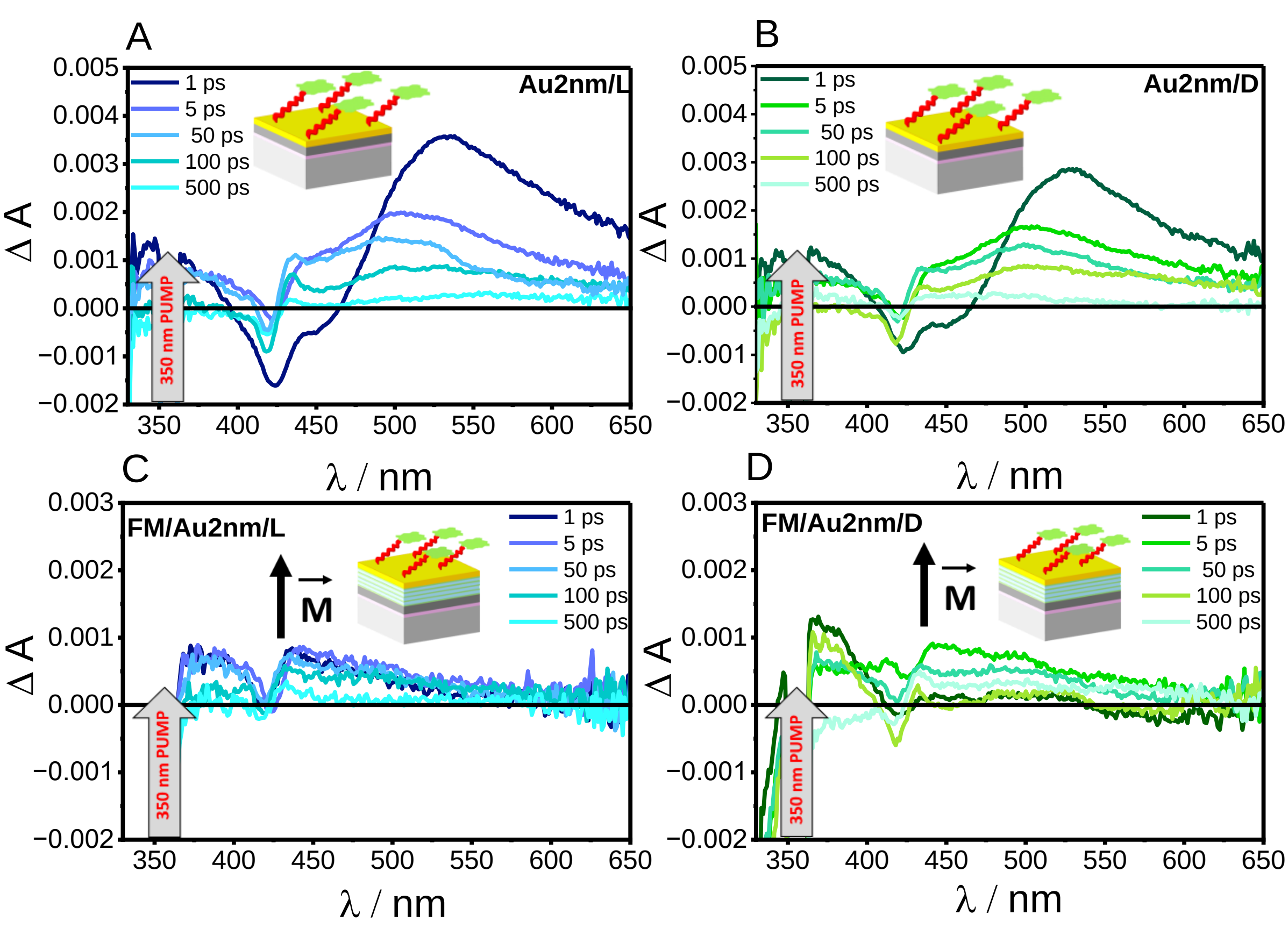

**Figure 8** Transient absorption spectra registered at various time delays for the following samples: A) 2 nm Au-L-porph, B) 2 nm Au-D-porph, C) FM/Au-L-porph (magnetization "up") and D) FM/Au-D-porph (magnetization "up") following the 350 nm excitation.

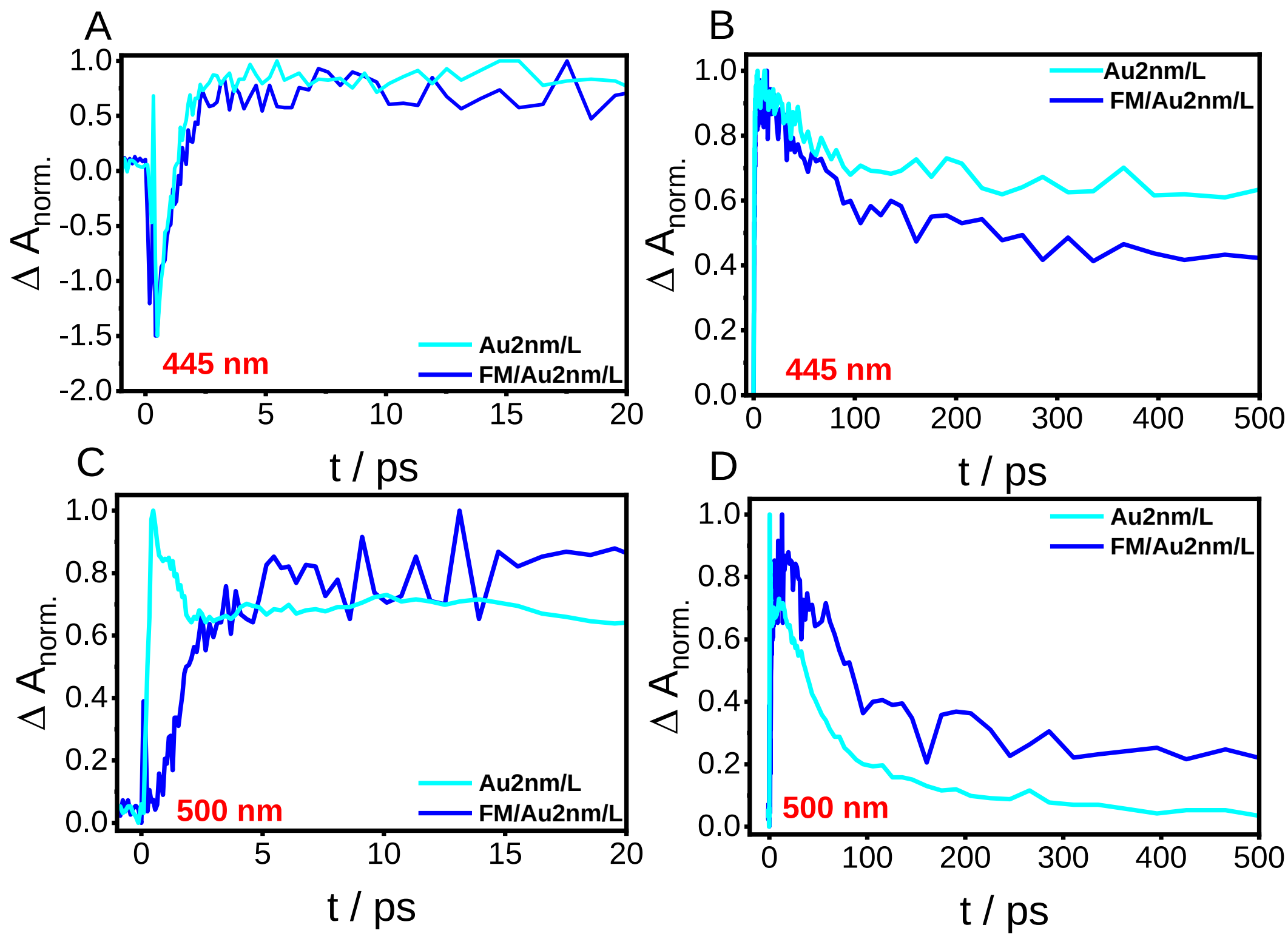

**Figure 9**. Transient absorption decays at 445 nm and 500 nm on short- and long-time scale for metallic heterostructures with L-porph anchored to the gold surface following the 350 nm laser excitation. Samples with a ferromagnetic layer were magnetized “up”.

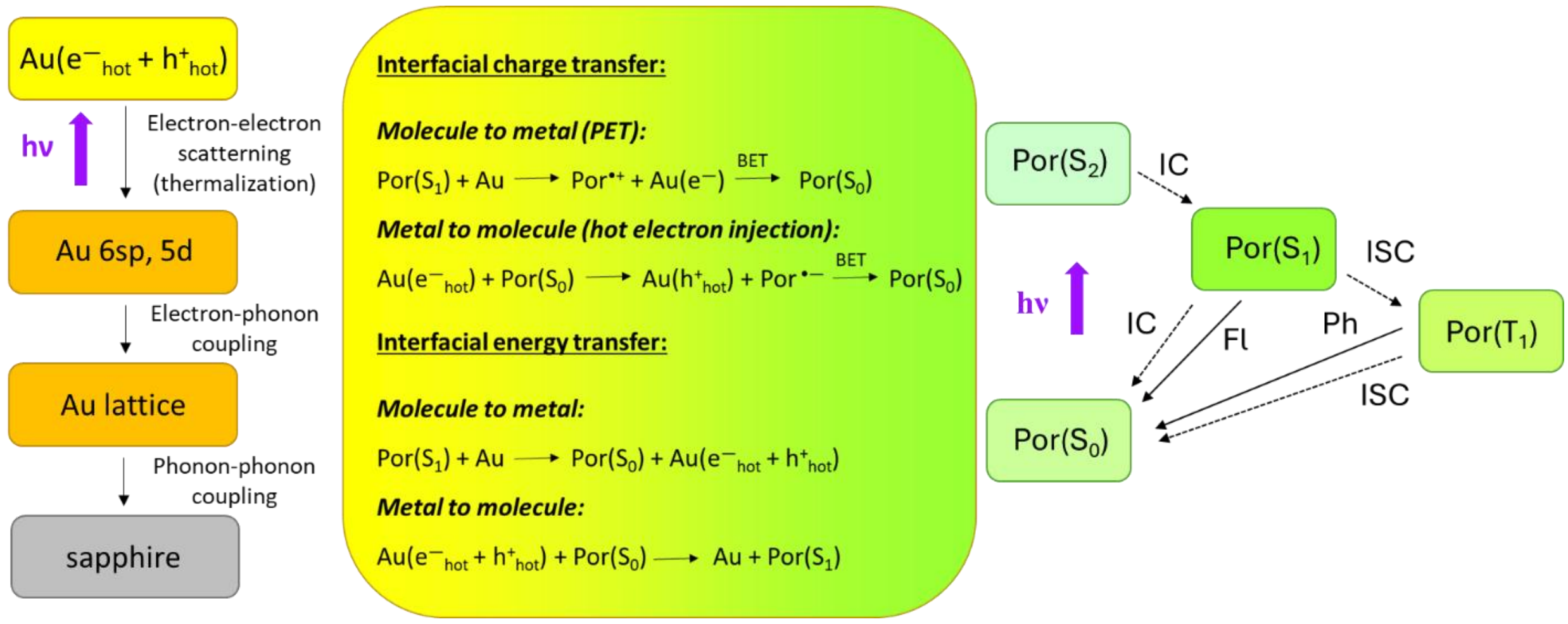

**Scheme 1** Schematic representation of the excitation pathways and possible relaxation processes in the studied hybrid systems. PET – photoinduced electron transfer, BET – back

electron transfer, IC – internal conversion, Fl – fluorescence, Ph -phosphorescence, ISC – intersystem crossing.

The results of our transient absorption measurements can be rationalized by the processes schematically depicted in Scheme 1. Upon photoexcitation, the excitation energy is predominantly absorbed by the metallic heterostructure, leading to the generation of hot electrons in the gold layer during the pump pulse. Nevertheless, direct excitation of the SAM also occurs to some extent, initially populating the $S_2$ excited state of the porphyrin chromophores, which rapidly relaxes via internal conversion to the $S_1$ state. In addition, indirect excitation of the chromophores via interfacial energy transfer from the excited metal occurs within the first ~ 2 ps: the rapid decay of hot electrons observed at 500 nm coincides with the rise of the positive signal at 445 nm, attributed to the porphyrin $S_1$ state.

Although the pump energy is predominantly absorbed by the metal stack, the transient absorption response is largely governed by the SAMs, except in Au/SAM samples excited at 350 nm. This is because the changes in the optical properties of the metal induced by hot electrons are comparatively small; thus, even strong absorption by the metal may result in a relatively weak spectroscopic signature. In the absence of the molecular layer, hot electrons in gold undergo ultrafast thermalization, followed by cooling through electron–phonon coupling and subsequent heat dissipation into the substrate.

In contrast, photoexcited porphyrins in solution relax via internal conversion, fluorescence, and intersystem crossing, exhibiting nanosecond $S_1$ excited-state lifetimes. Upon immobilization on metallic heterostructures, however, the excited-state lifetime is drastically reduced (< 50 ps).[59,97,98] This quenching is attributed to the proximity of the gold surface, which provides an efficiently competing nonradiative relaxation pathway via electron–hole pair generation in the metal (i.e., interfacial energy transfer from the molecule to the metal), thereby significantly accelerating excited-state decay.[97–99] Additionally, interactions between closely packed chromophores within the SAM may further shorten the excited-state lifetime via self-quenching.[59,91] Therefore, the fast decay component at 445 nm can be attributed to the relaxation of the porphyrin $S_1$ state via energy transfer to the metal and possibly self-quenching in SAM. The slow decay component, on the other hand, can be assigned to the heat dissipation from the Au lattice to the substrate. At the same time, the slow rise of the positive signal observed at 500 nm around 10-20 ps can partially be related to the mentioned energy transfer from the porphyrin to the metal, which occurs on a comparable time scale.

Although interfacial electron transfer in the studied systems is thermodynamically feasible, no evidence for the formation of porphyrin radical cations or anions was observed. This can be rationalized in two ways. First, ultrafast back electron transfer may suppress the accumulation of charge-transfer states, as is commonly observed in similar systems.[100–102] Second, highly efficient energy transfer from the excited molecules to the metal may compete with, and even dominate, the electron transfer pathway.[98]

Several structural modifications could be envisaged to promote electron transfer from the chromophore to the metal. These include metallation of the porphyrin macrocycle to enhance the ease of its oxidation,[103] incorporation of electron-donating substituents to stabilize the charge transfer products, and attachment of the chromophore at the C-terminus of the oligopeptide while anchoring the molecule to the metal via a cysteine residue at the N-terminus. The latter configuration would favor electron transfer through the helix, as its macrodipole could stabilize the charge-separated state.[22,37,40] Furthermore, shortening the oligopeptide linker would increase electronic coupling between the metal surface and the porphyrin macrocycles, which is beneficial for electron transfer, though it may also enhance competing energy transfer processes.[80]

Finally, the absence of any measurable dependence of the photoinduced dynamics on the chirality of the molecules or on the direction of the ferromagnet magnetization can most likely be attributed to the lack of efficient interfacial electron transfer. Since spin-dependent effects are expected to primarily influence electron transfer processes, their absence suggests that these processes do not play a dominant role in the relaxation dynamics under the present conditions.

## 3 Conclusion

In this work, we investigated the ground- and excited-state properties of chiral porphyrin self-assembled monolayers (SAMs) on gold-capped metallic heterostructures with and without an underlying ferromagnetic layer. Steady-state absorption measurements show that SAM formation leads to a red shift and broadening of the Soret band, which we attribute to head-to-tail π–π interactions between neighboring porphyrin macrocycles within the monolayer. These spectral features, together with the estimated surface coverages, indicate a predominantly parallel orientation of the chromophores relative to the gold surface.

Despite the dominant absorption of the pump pulse by the metallic stack, excitation of the porphyrin SAM either via direct absorption of light and/or indirectly via energy transfer from the metallic heterostructure, is clearly observed and gives rise to characteristic ground-state bleach and excited-state absorption features. The porphyrin $S_1$ excited state decays on a sub-nanosecond timescale when adsorbed on the metallic heterostructure, in contrast to its nanosecond lifetime in solution, indicating efficient interfacial quenching processes. However, no clear signatures of long-lived charge-transfer states were detected.

Importantly, neither the chirality of the SAM nor the magnetization direction of the ferromagnetic layer has a measurable effect on the transient absorption spectra or kinetics under the investigated conditions. This suggests that, although spin polarization at the interface and chiral-induced spin selectivity may be present, they do not significantly influence the dominant relaxation pathways, which are governed by ultrafast energy dissipation across the interface.

Overall, our results highlight the critical roles of interfacial energy transfer and hot-electron dynamics in determining the photophysics of molecule–metal hybrid systems and demonstrate that efficient photoinduced electron transfer in such architectures requires not only favorable energetics but also sufficiently strong electronic coupling and optimized interfacial structure.

## 4. Experimental Section

*Materials:* 5-(4-carboxyphenyl)-10,15,20-(triphenyl)porphyrin (TPCOOH) was purchased from PorphyChem. Synthesis of the chiral oligopeptide linker composed of alternating alanine (Ala) and 2-aminoisobutyric acid (Aib) residues with C-terminal cysteine $(Ala\text{-}Aib)_8$-Cys-OH as well as N-terminal modification of this oligopeptide with TPCOOH chromophore yielding L-porph and D-porph derivatives was carried out by the Peptide Specialty Laboratories GmbH. Tetraphenylporphyrin (TPP) and ethanol (EtOH, HPLC grade) were purchased from Sigma-Aldrich and used with no further purifications.

*Preparation of metallic Ta/Pt/[Co/Ni]$_4$/Co/Au films:* To investigate the impact of the magnetization of FM on the spectroscopic properties of porphyrin molecules covalently bound to FM/Au via chiral linkers, we focus on magnetic multilayers that are characterized by strong perpendicular magnetic anisotropy (PMA), ensuring that the FM magnetization remains perpendicular to the film plane in the ground state also in the absence of an external magnetic field. Co/Ni multilayers, in particular, are identified as optimal systems due to their pronounced PMA, which results in a high coercive field. In addition, they demonstrate an elevated Curie temperature and saturation magnetization, indicating a larger energy splitting between the majority and minority spin bands along with significant spin polarization. These essential properties of Co/Ni multilayers are utilized in numerous spintronic applications.[104–108]

The Co/Ni multilayers were fabricated using direct current (dc) magnetron sputter deposition in an argon atmosphere at a pressure of $3\times10^{-3}$ mbar within an ultrahigh-vacuum BESTEC system, which has a base pressure of $5\times10^{-9}$ mbar. C-plane sapphire substrates ($Al_2O_3$ (0001)), each with a thickness of 0.5 mm and a size of $0.5\times0.5$ mm$^2$, were used for film growth. A total of 10 substrates were placed on a sample holder, enabling uniform, controlled film growth across all substrates. Prior to the deposition of the multilayers, a 1.5 nm Ta layer was applied to enhance adhesion. A subsequent 3-nm-thick Pt layer was deposited to promote a preferred Co/Ni fcc (111) texture, which contributes to improved growth and larger PMA.[105,108] The multilayers were capped with a 2 nm Au layer to prevent surface oxidation and to enable covalent attachment of L-porph and D-porph.

All metallic layers were deposited at room temperature using alternating source shutters. To ensure uniform growth of the entire stack, the sample holder was rotated at approximately 30 rpm. The target guns for Co and Ni, were positioned at an angle of 30° relative to the substrate normal, while the Ta, Pt and Au layers were deposited in a face-to-face configuration to achieve lower sputter rates, controlled by the target-to-substrate distance, thus ensuring homogeneous coating. Additionally, to further promote uniform growth, the dc power for the Pt target was reduced to 40 W, while Au was deposited using radio frequency (rf) sputtering at 30 W. The thickness of each layer was controlled via deposition time. Prior to sample fabrication, the sputter rates were calibrated using X-ray reflectivity, yielding rates of 0.045 nm/s for Co at 50 W dc power, 0.068 nm/s for Ni at 60 W, 0.06 nm/s for Pt at 40 W, and 0.07 nm/s for Au at 30 W rf power.

Following the optimization of magnetic properties, **two sets of metallic heterostructures** were prepared for this study, featuring the following structure:

1) $Al_2O_3$(sub.)/Ta(1.5nm)/Pt(3nm)/[Co(0.25nm)/Ni(0.65nm)]$_3$/Co(0.25nm)/Au(2nm) - abbreviated as **FM/Au**

2) $Al_2O_3$(sub.)/Ta(1.5nm)/Pt(3nm)/Au(2nm) - abbreviated as **Au**

The second set of samples serves as a reference for studying the spectroscopic properties of porphyrin molecules adsorbed on the Au surface in the absence of the ferromagnetic layer. Comparing these samples with the magnetic ones allows us to isolate the effect of the spin-split band structure induced by the ferromagnetic Co/Ni monolayers.

*Self-assembly of chiral porphyrins:* Prior to the self-assembly of the molecules, all samples were thoroughly cleaned with organic solvents: 15 minutes of soaking in acetone, 15 minutes in isopropanol, and 15 minutes in ethanol followed by $N_2$-blow dry. The whole self-assembly was carried out in a glove box under an Ar atmosphere. Ethanolic solutions of porphyrin derivatives were prepared via solubilization of ca. 0.5-0.6 mg of the given compound in 3 mL of absolute ethanol, which corresponds to ca. 90 µM concentration. As discussed in paragraph *Photophysical properties of chiral porphyrins in solution*, this concentration is low enough to ensure the presence of porphyrin derivatives in their monomeric (non-aggregated) form. To ensure identical concentrations of the L-porph and D-porph solutions, their absorption spectra were measured, and the solutions were subsequently diluted to match their absorbances at the Soret band (415 nm). In the next step, the FM/Au and Au thin films were immersed in 400 µL of the solution for 24 hours. After that time the samples were taken out from the solution, rinsed with ethanol to remove any unbound porphyrin derivative and $N_2$-blow dried. As fabricated organic/inorganic hybrids were stored in dark and under vacuum. Reference samples were prepared in an identical manner, using pure ethanol instead of the ethanolic solution of the porphyrin derivative.

*Characterization:* Cary 500 UV-Vis two-beam spectrometer was used to record UV-Vis absorption spectra in the range from 800 to 200 nm with 1 nm step, using quartz cells for solutions with various optical path length (2–10 mm) or holder for solid state samples. Fluorescence spectra were recorded in the range of 630–800 nm on a JASCO FP-8550 spectrofluorometer for diluted solutions with an absorbance at the excitation wavelength lower than 0.1. Fluorescence quantum yield ($\phi_F$) was determined using a standard substance. TPP has been used as standard in experiments to determine fluorescence quantum yield.[109] The fluorescence lifetimes were measured on a Fluorescence Lifetime Spectrometer (FluoTime300 from PicoQuant) using a time-correlated single-photon counting (TCSPC) detection system. The emission decay lifetimes were acquired using 405 nm diode laser as the excitation source.

In addition, an instrument response function (IRF) was obtained using Ludox solution (colloidal silica). Circular dichroism spectra of chiral porphyrins in degassed ethanolic solution together with spectra of the oligopeptides without chromophores, were registered using a JASCO J-810 spectropolarimeter. ECD spectra of the analytes were measured with 4 accumulations, using a quartz cell with an optical length of 1 mm.

The femtosecond transient absorption spectroscopy setup consisted of a short-pulse titanium sapphire oscillator (Mai Tai, Spectra Physics, 70 fs) followed by a high-energy titanium sapphire regenerative amplifier (Spitfire Ace, Spectra Physics, 100 fs). The 800 nm beam was split into two beams to generate: (1) a pump ($\lambda_{exc}$ = 390 nm for solutions, $\lambda_{exc}$ = 350 nm and 420 nm for solid samples) from the optical parametric amplifier (Topas Prime with a NirVis frequency mixer) and (2) probe pulses in the UV-Vis range by using $CaF_2$ (Ultrafast Systems, Helios). The temporal resolution of the setup is about 200 fs. Porphyrin solutions were prepared by dissolving the solid samples in EtOH to achieve a concentration of approximately $10^{-5}$ M. The stationary absorption spectra were recorded to verify the presence of a well-defined Soret band, ensuring adequate sample quality. All measurements in solutions were performed using a 2 mm quartz cuvette, with continuous stirring via a magnetic stirrer to maintain homogeneity. To confirm the stability and integrity of the sample, UV-Vis spectra were collected before and after each experiment. Typical pump pulse energy was about 0.2 µJ (the pump diameter FWHM ≈ 200-250 µm). All experiments were performed at room temperature. The solid sample was mounted so that the pump pulse and probe pulse overlapped. To ensure sample stability, the transient spectra were recorded for only 144 various time delays. The transient absorption spectra for the solid sample were measured twice by changing the magnetization orientation between the measurements. Analysis of the transient absorption data was made using Surface Explorer software (Ultrafast Systems). The acquired experimental data were analyzed using the commercial program OriginPro 2023b.

**Acknowledgements**

This research was financially supported by the National Science Centre (project no. 2020/39/I/ST5/00597) and the German Research Foundation (DFG, Deutsche Forschungsgemeinschaft, project no. 464974971).

**Conflict of Interest**

The authors declare no conflict of interest

**Data Availability Statement**

The data that support the findings of this study are available from the corresponding author upon reasonable request.

# Supporting Information

Chiral Porphyrin Monolayers on Ferromagnetic Thin Films: Ultrafast Spectroscopy of Hybrid Interfaces

*Karol Hauza, Anna Lewandowska-Andralojc*[*]*, Ruslan Salikhov, Jürgen Lindner, Gotard Burdzinski, Marcin Kwit, Bronislaw Marciniak and Aleksandra Lindner*[*]

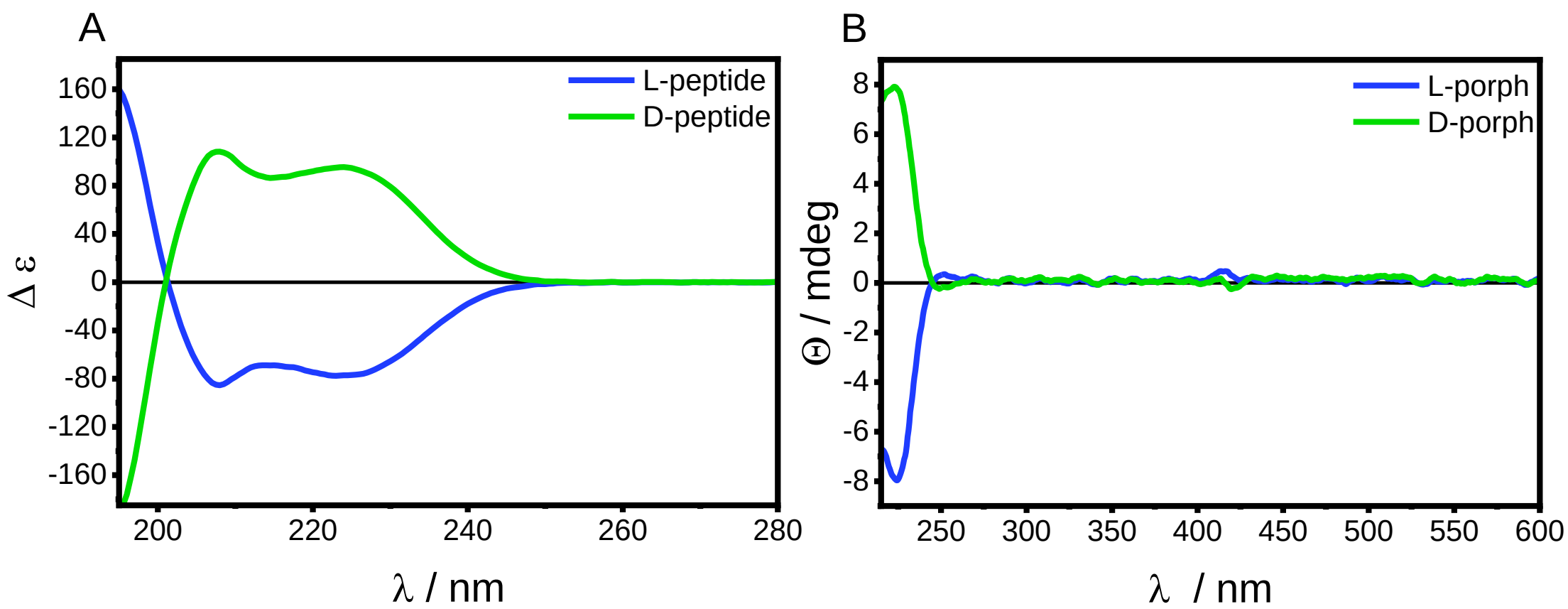

**Figure S1.** Circular dichroism spectra of: A) chiral peptides without porphyrin chromophore and B) the same chiral peptides with porphyrin chromophore in EtOH.

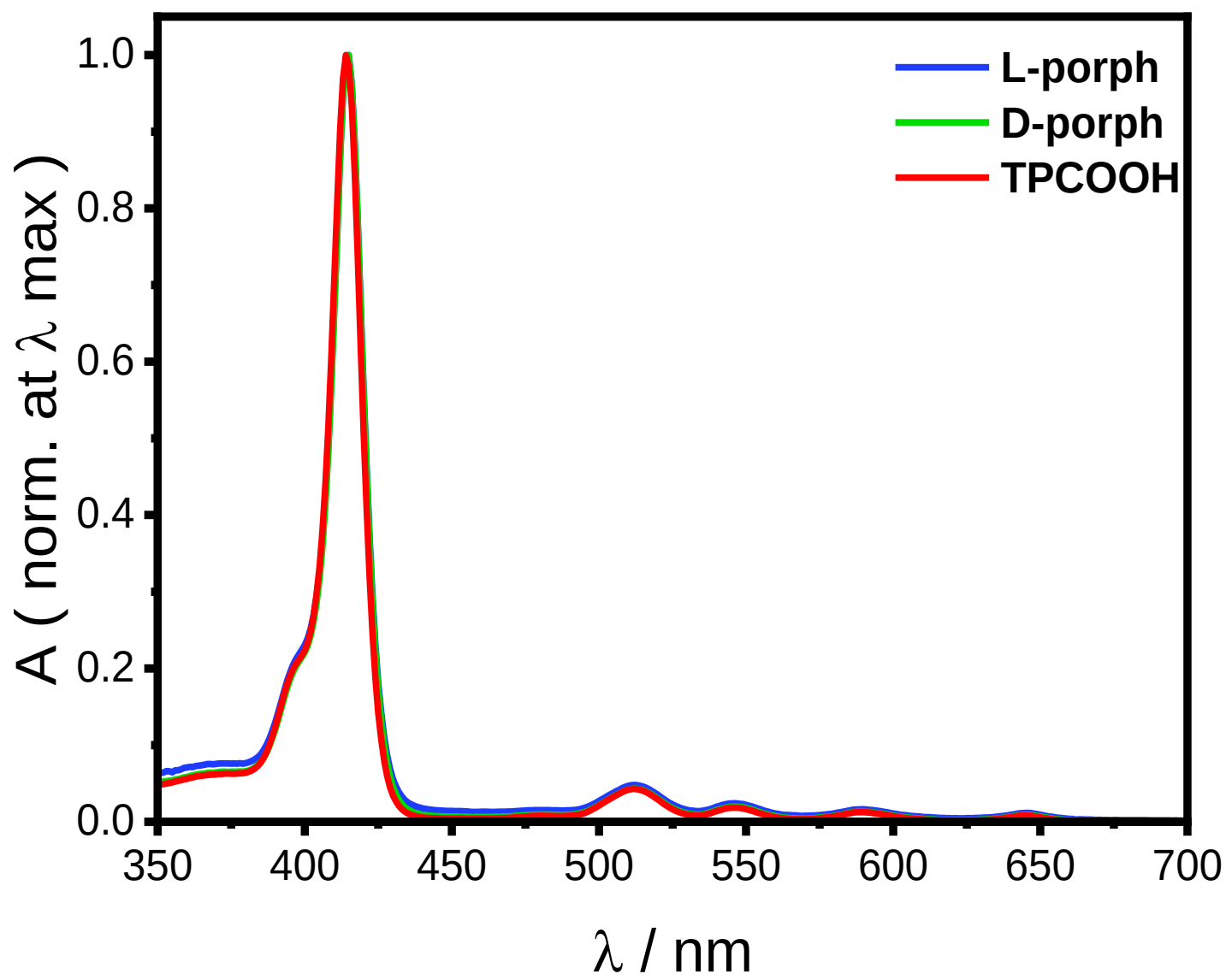

**Figure S2.** Normalized absorption spectra of L-porph, D-porph and TPCOOH registered in EtOH.

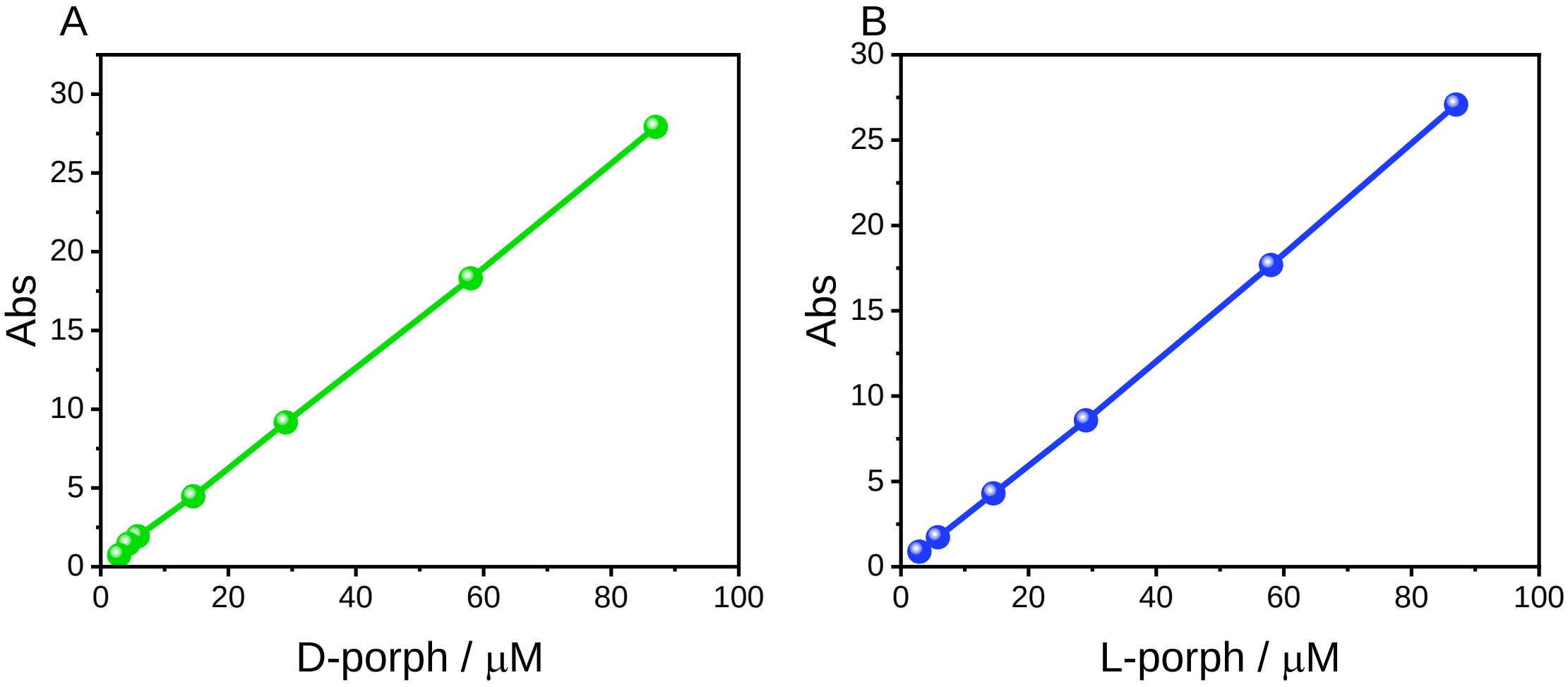

**Figure S3** Dependence of the absorbance at the 415 nm on the concentration of chiral porphyrin: A) D-porph and B) L-porph. NOTE: The absorption spectra with absorbance maxima exceeding 2, were actually obtained by recalculating the absorbance values for samples measured in cuvettes with an optical path length of 0.2 cm or 0.1 cm in accordance with the Lambert-Beer law.

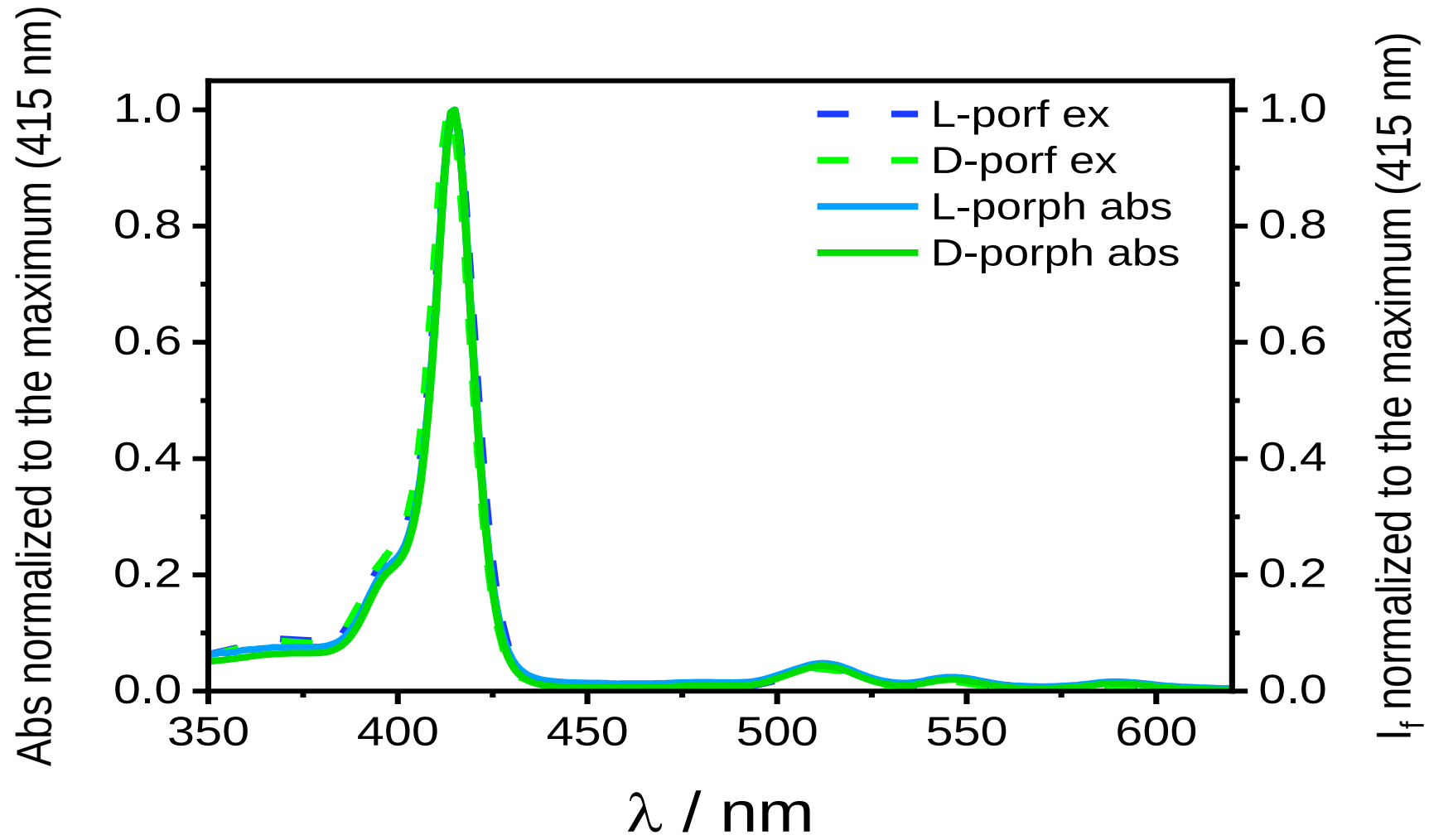

**Figure S4.** Comparison of the normalized absorption and excitation spectra of chiral porphyrins in ethanol ($\lambda_{em}$= 647 nm, excitation and emission slits 2.5 nm).

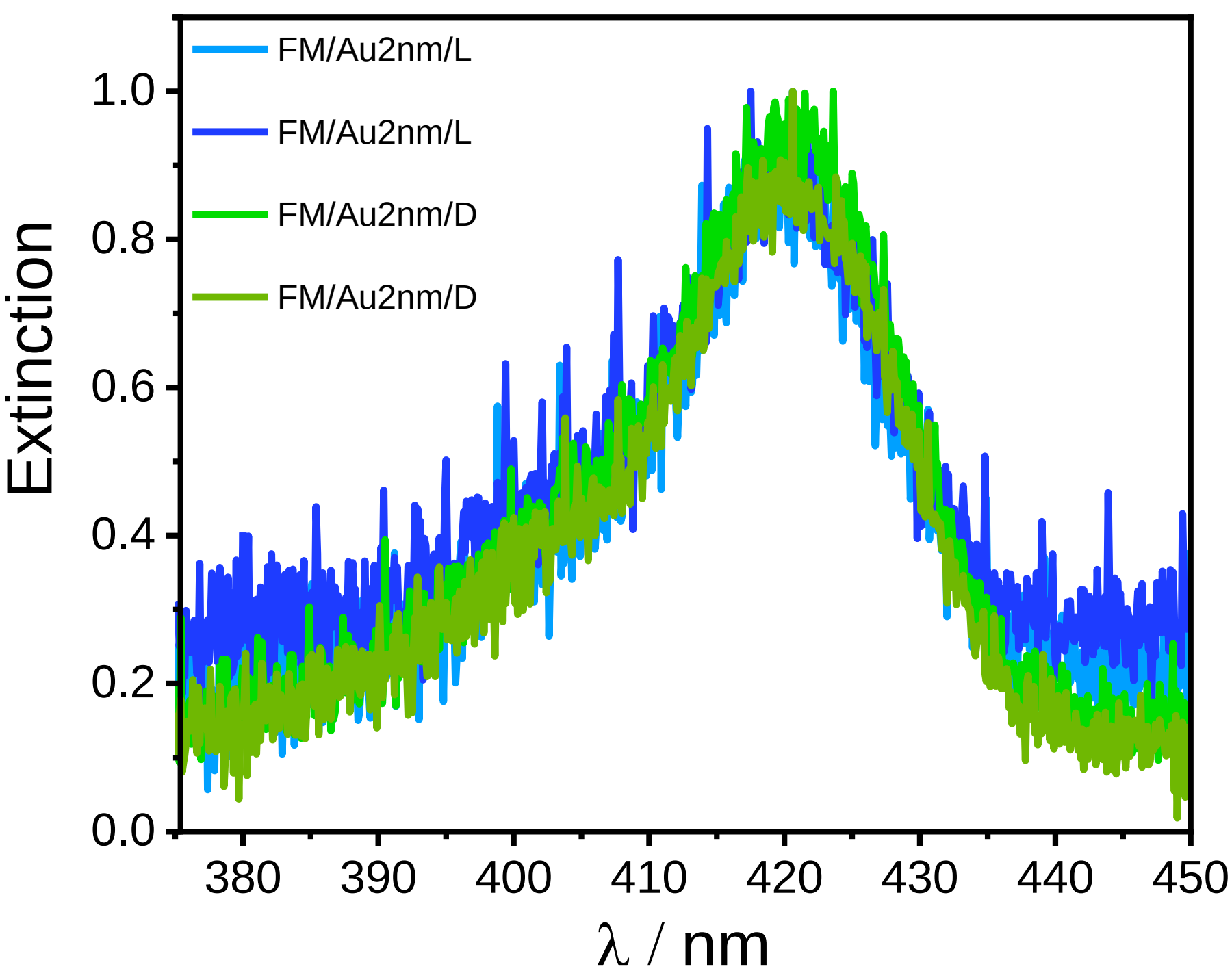

**Figure S5.** Comparison of the normalized extinction spectra of FM/Au- L-porph and FM/Au-D-porph for two opposite directions of the out-of-plane magnetization polarization ("up" and "down").

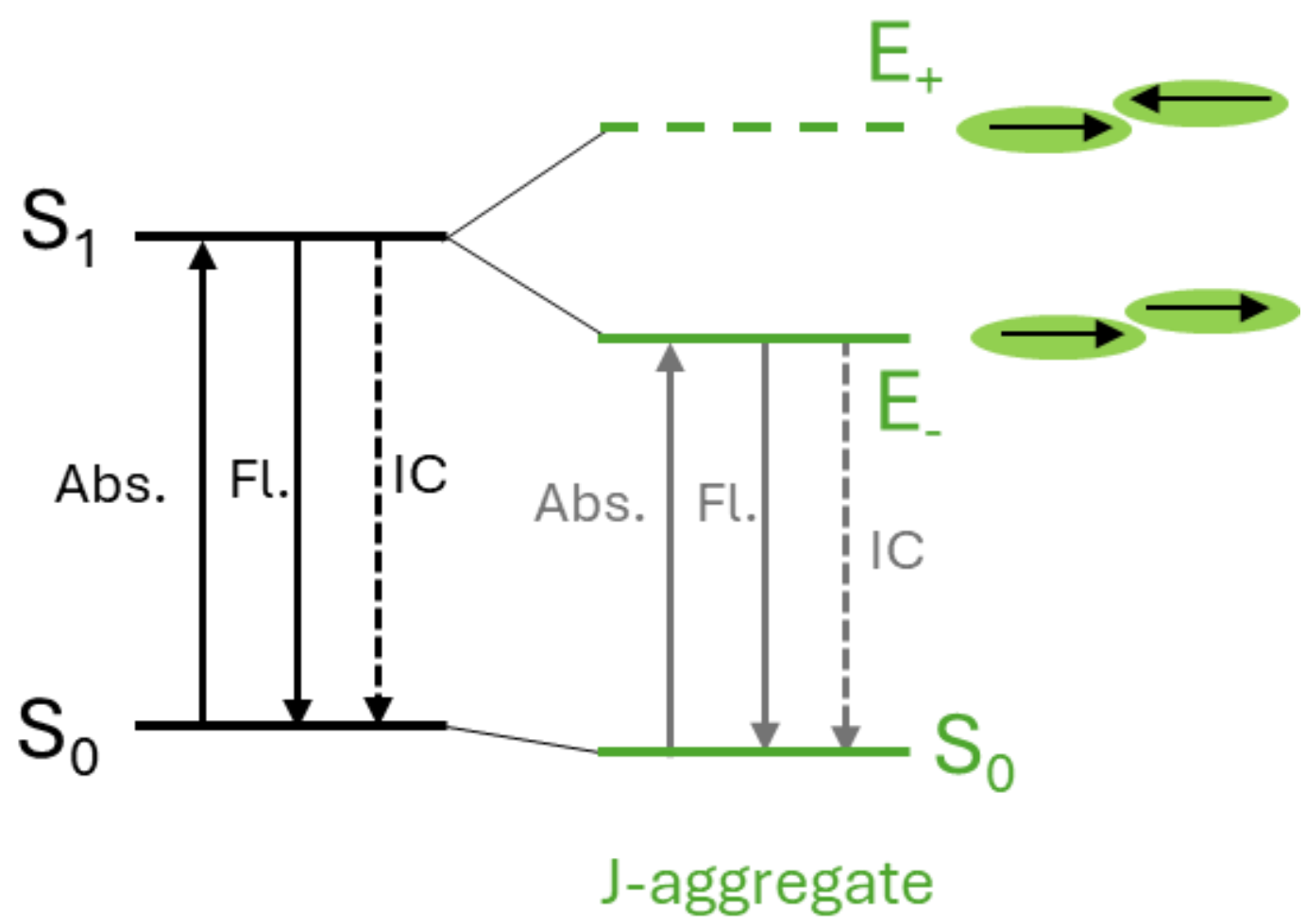

**Scheme S1.** Exciton bands energy diagram for the J-type porphyrin dimers (depicted without oligopeptide linker). Abs. – absorption, Fl. – fluorescence, IC – internal conversion, $E_+$ and $E_-$ – excitonic energy levels. The upper exciton state (dashed green line) with antiparallel alignment of transition dipole moments (black arrows) in porphyrin chromophores (light green oval forms) has a forbidden transition to the molecular ground state.

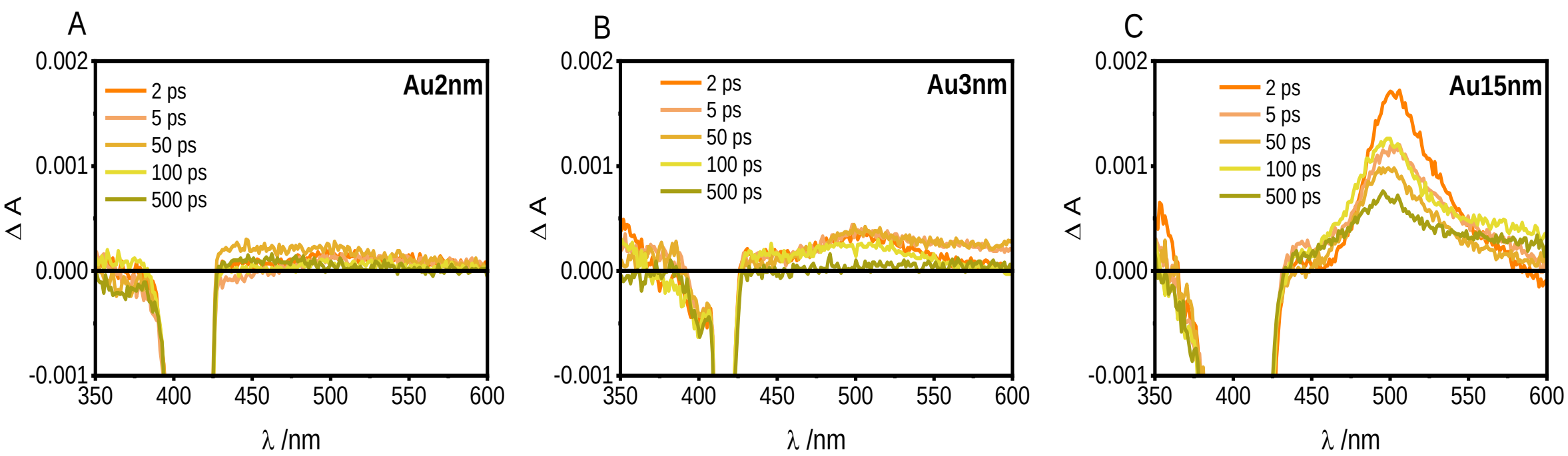

**Figure S6.** Transient absorption spectra registered at various time delays for Au reference samples with varying thickness of Au layer A) 2 nm, B) 3 nm C) 15 nm, following the 420 nm excitation.

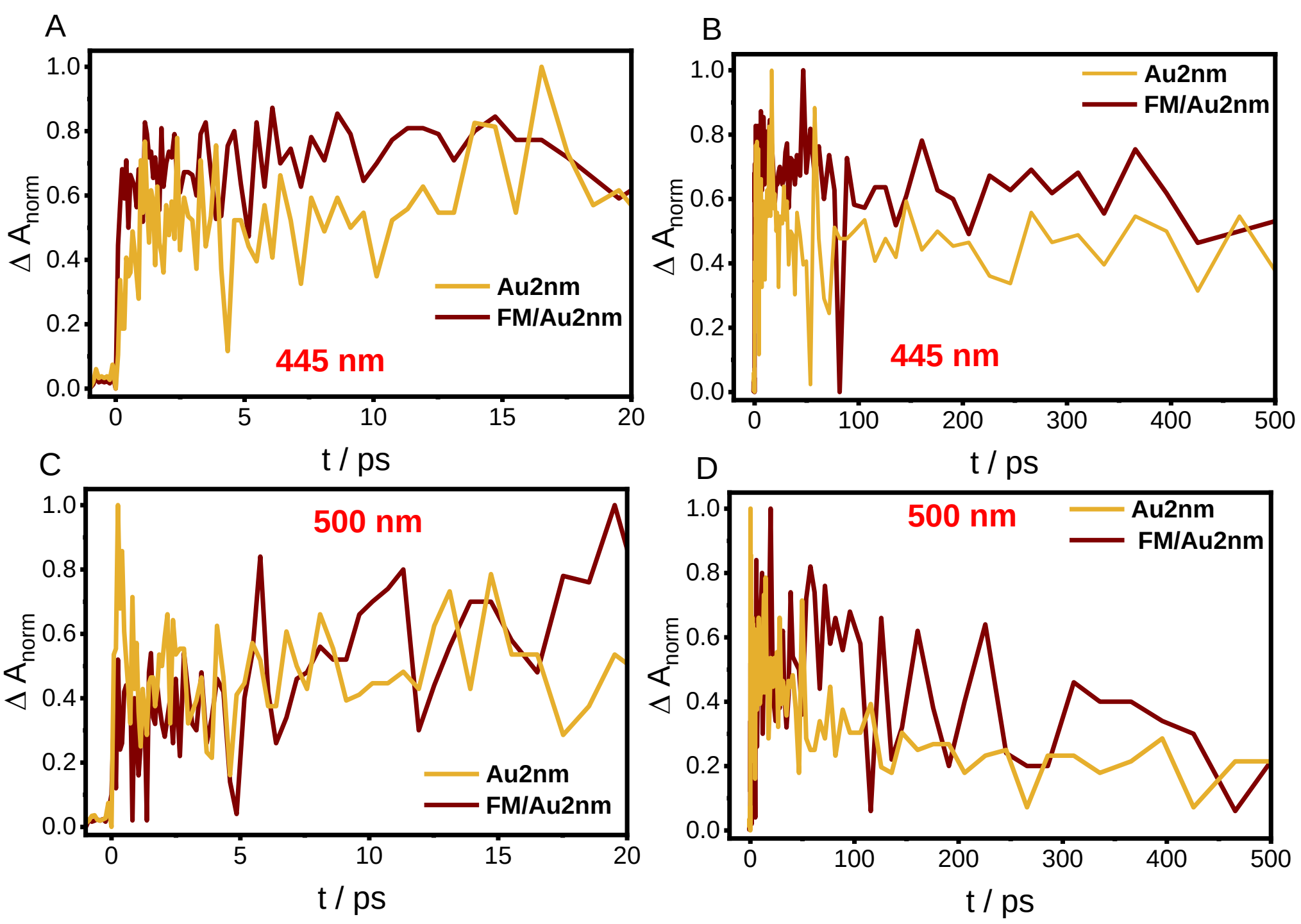

**Figure S7** Transient absorption kinetics taken at 445 nm and 500 nm following excitation with 420 nm pump for two reference samples: 2 nm Au and FM/Au 2nm.

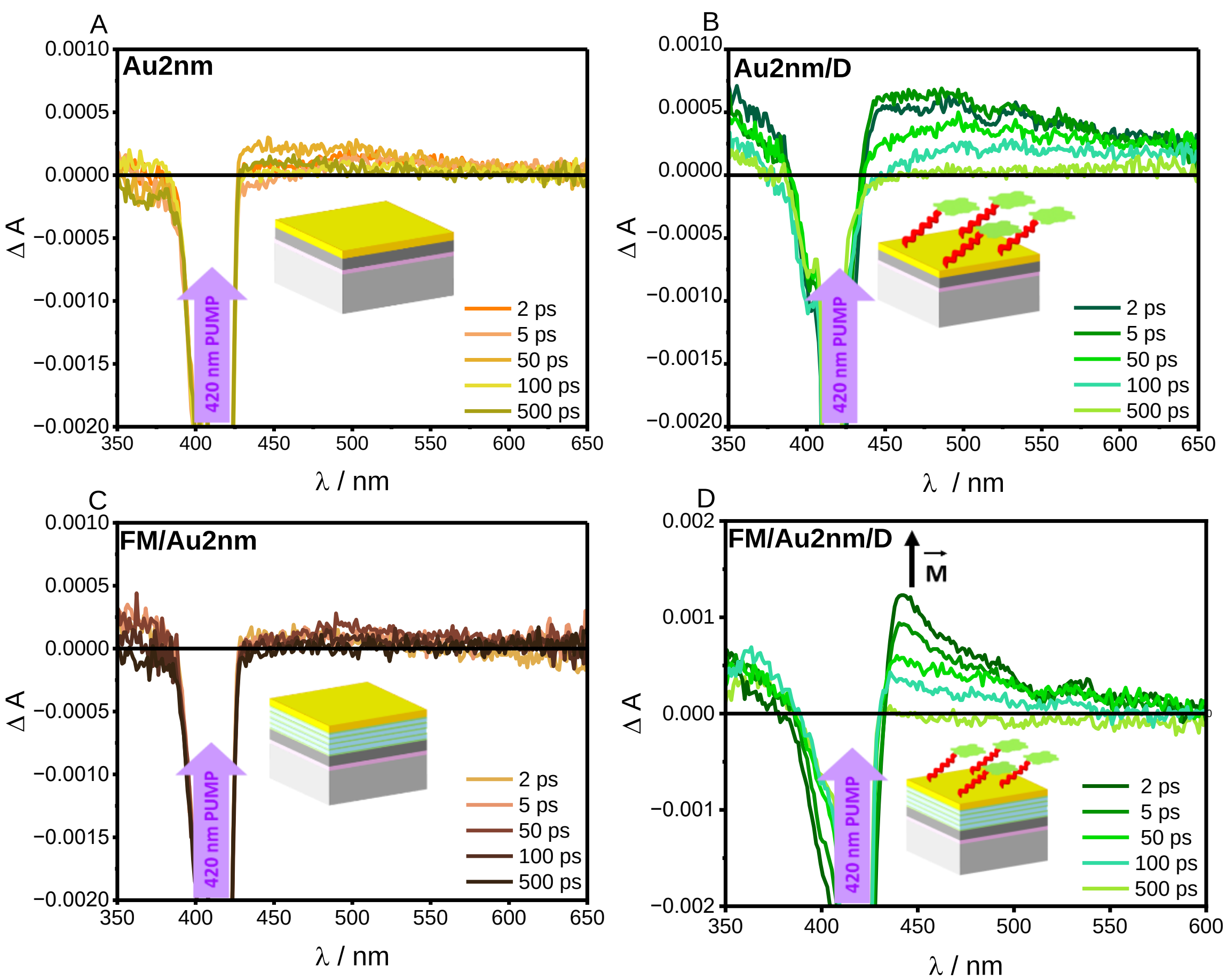

**Figure S8** Transient absorption spectra registered at various time delays for two samples without and with FM layer, both with SAMs of D-chirality: A) Au-D-porph and B) FM/Au-D-porph with magnetization "up" following the 420 nm excitation.

**Table S1** Fraction of the absorbed incident light together with fraction of the absorbed pump energy by SAM and metal stack for Au and FM/Au metallic heterostructures and SAM of D-chirality.

| Pump wavelength | Sample | Extinction at the pump wavelength | Absorber | Fraction of the absorbed incident light | Fraction of the absorbed pump energy |
|---|---|---|---|---|---|
| **350 nm** | S/Ta/Pt/**Au/SAM** | 0.428 (without SAM: 0.422) | Metal stack | 62.1% | 99% |
| | | | SAM | 0.6% | 1% |
| | S/Ta/Pt/**(Co/Ni)/Au/SAM** | 0.556 (without SAM: 0.553) | Metal stack | 72.0% | 99.7% |
| | | | SAM | 0.2% | 0.3% |
| **420 nm** | S/Ta/Pt/**Au/SAM** | 0.466 (without SAM: 0.423) | Metal stack | 62.2% | 95% |
| | | | SAM | 3.5% | 5% |
| | S/Ta/Pt/**(Co/Ni)/Au/SAM** | 0.604 (without SAM: 0.558) | Metal stack | 72.3% | 96.3% |
| | | | SAM | 2.8% | 3.7% |

Results summarized in Table S1 were obtained from the extinction spectra of various samples in a manner analogous to the one presented below for the 350 nm excitation of the Au/SAM:

**Absorbance of the metal stack:**

$A_{metal} = 0.422 \Rightarrow T = 0.379$

100% - 37.9% = 62.1% => 62.1% of the incoming light is absorbed by the metal stack

**Absorbance of a metal stack after SAM adsorption:**

$A_{metal+SAM} = 0.428$

$$T_{metal+SAM} = 10^{-A_{metal+SAM}} = 0.373$$

**Decrease of the transmission due to light absorption in SAM:**

$\Delta T = T_{metal} - T_{metal+SAM} = 0.379 - 0.373 = 0.006$ => 0.6% of the incoming light is absorbed by the SAM

**The excitation partition** (fraction of the absorbed pump energy by each of the components)

$$\left(\frac{\%\ of\ light\ absorbed\ by\ a\ component}{\%\ of\ light\ absorbed\ by\ all\ the\ components}\right) \times 100\%$$

**Example:** Fraction of the pump energy absorbed by SAM in S/Ta/Pt/Au/SAM when excited with 350 nm pump.

$$\left(\frac{0.6\%}{0.6\% + 62.1\%}\right) \times 100\% = 0.95\% \sim 1\%$$

**1%** of the pump energy is absorbed by **SAM**

**99%** of the pump energy is absorbed by the **metal stack**

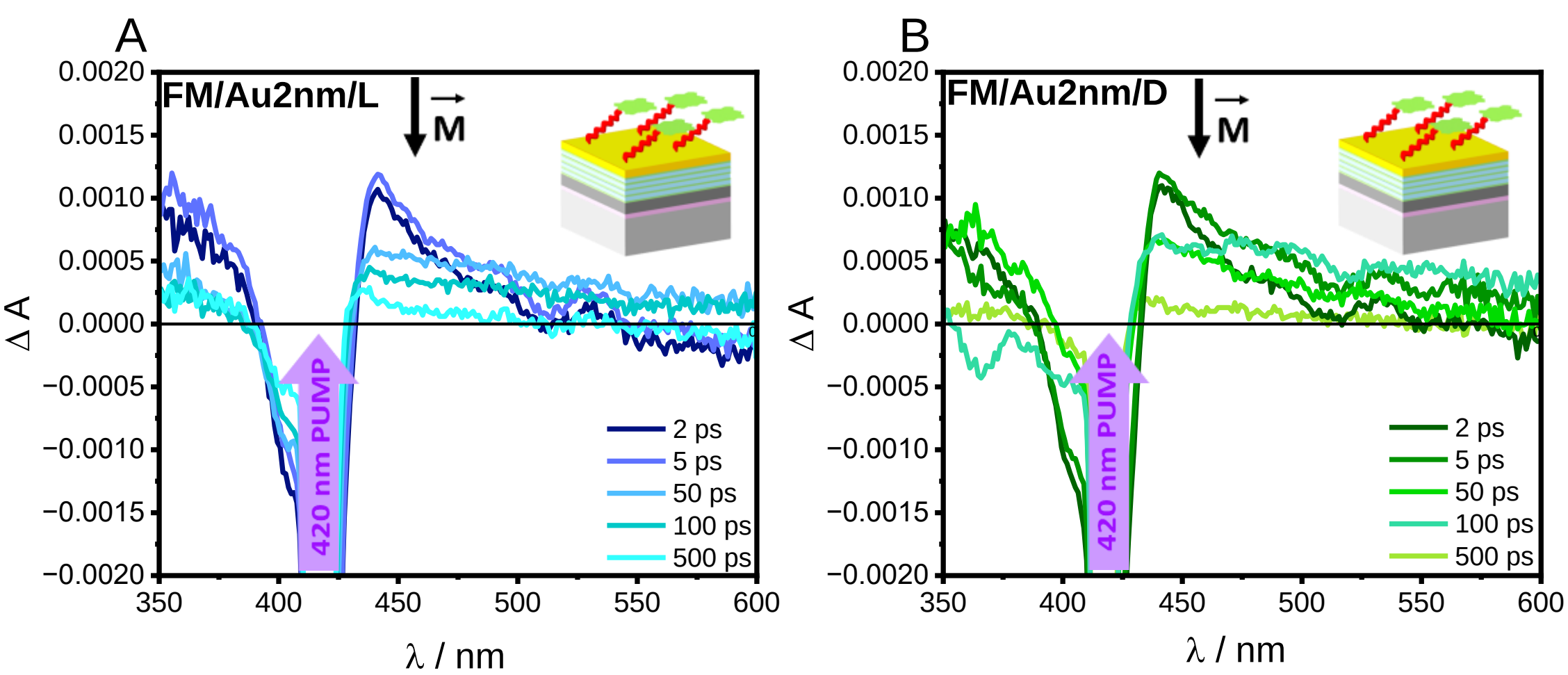

**Figure S9** Transient absorption spectra registered at various time delays for two samples with FM layer magnetized “down”: A) with L-porph SAM and B) with D-porph SAM, following the 420 nm excitation.

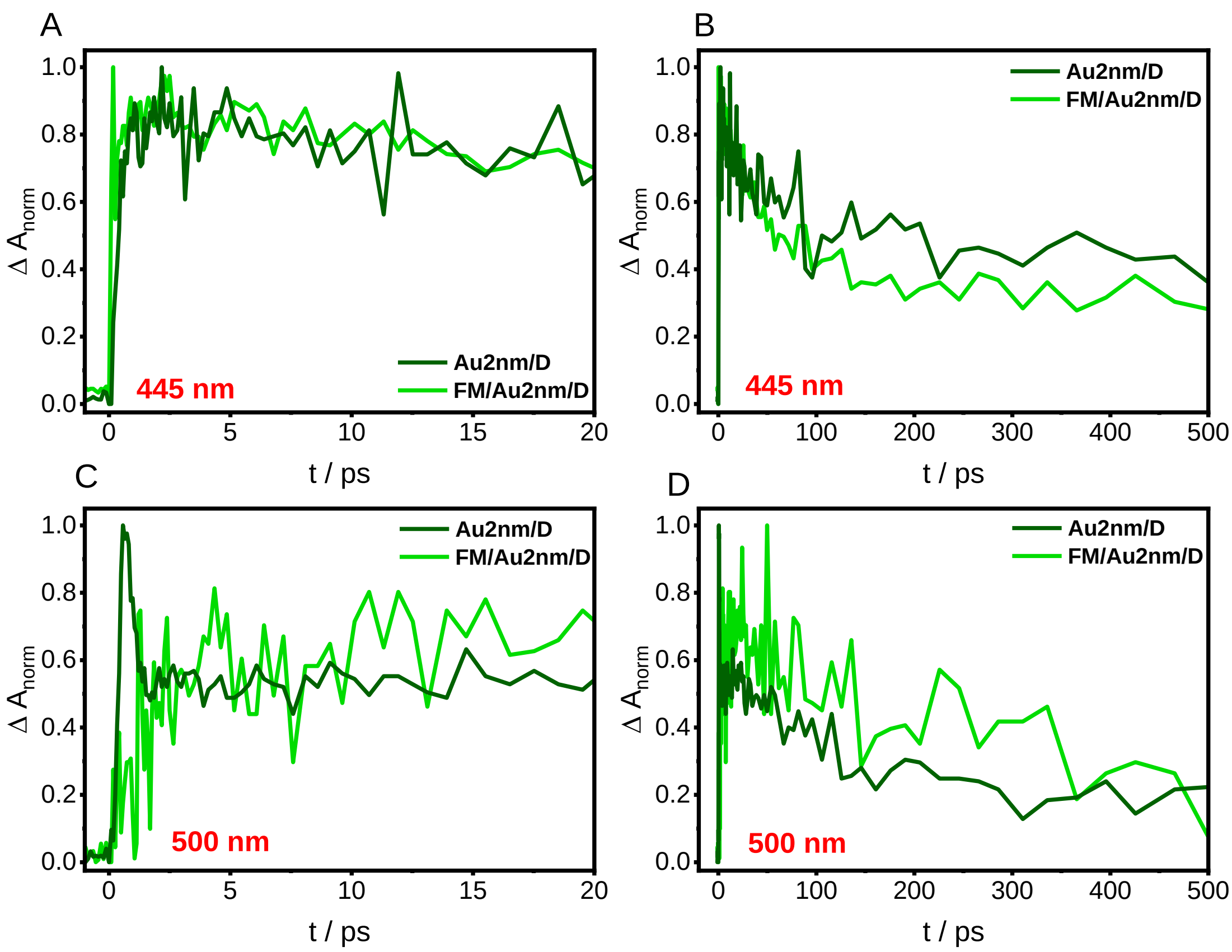

**Figure S10** Transient absorption kinetics at 445 nm and 500 nm on short- and long-time scale for metallic heterostructures with D-porph anchored to the gold surface following the 420 nm laser excitation. FM magnetization pointing "up". NOTE: In panel D, a distinct sharp feature is observed at approximately 50 ps. After normalization, this peak is set to 1, which consequently makes the signal in panel B appear lower in comparison to Au2nm/D.

**Table S2** Compilation of the time constants obtained via curve fitting at 445 nm with biexponential function for various metallic heterostructures without and with chiral molecules attached to the gold surface following excitation with 420 nm.

| Sample | Decay kinetics at 445 nm | |
|---|---|---|
| Au | [a] | |
| Au/L | $t_1$ = 22 ps, $t_2$ = 203 ps | |
| Au/D | $t_1$ = 29 ps, $t_2$ = 219 ps | |
| | Magnetization UP | Magnetization DOWN |
| FM/Au | [a] | [a] |
| FM/Au/L | $t_1$ = 36 ps, $t_2$ =234 ps | $t_1$ = 39 ps, $t_2$ = 259 ps |
| FM/Au/D | $t_1$ = 24 ps, $t_2$ = 256 ps | $t_1$ = 21 ps, $t_2$ = 218 ps |

a- cannot be determined due to the low signal intensity

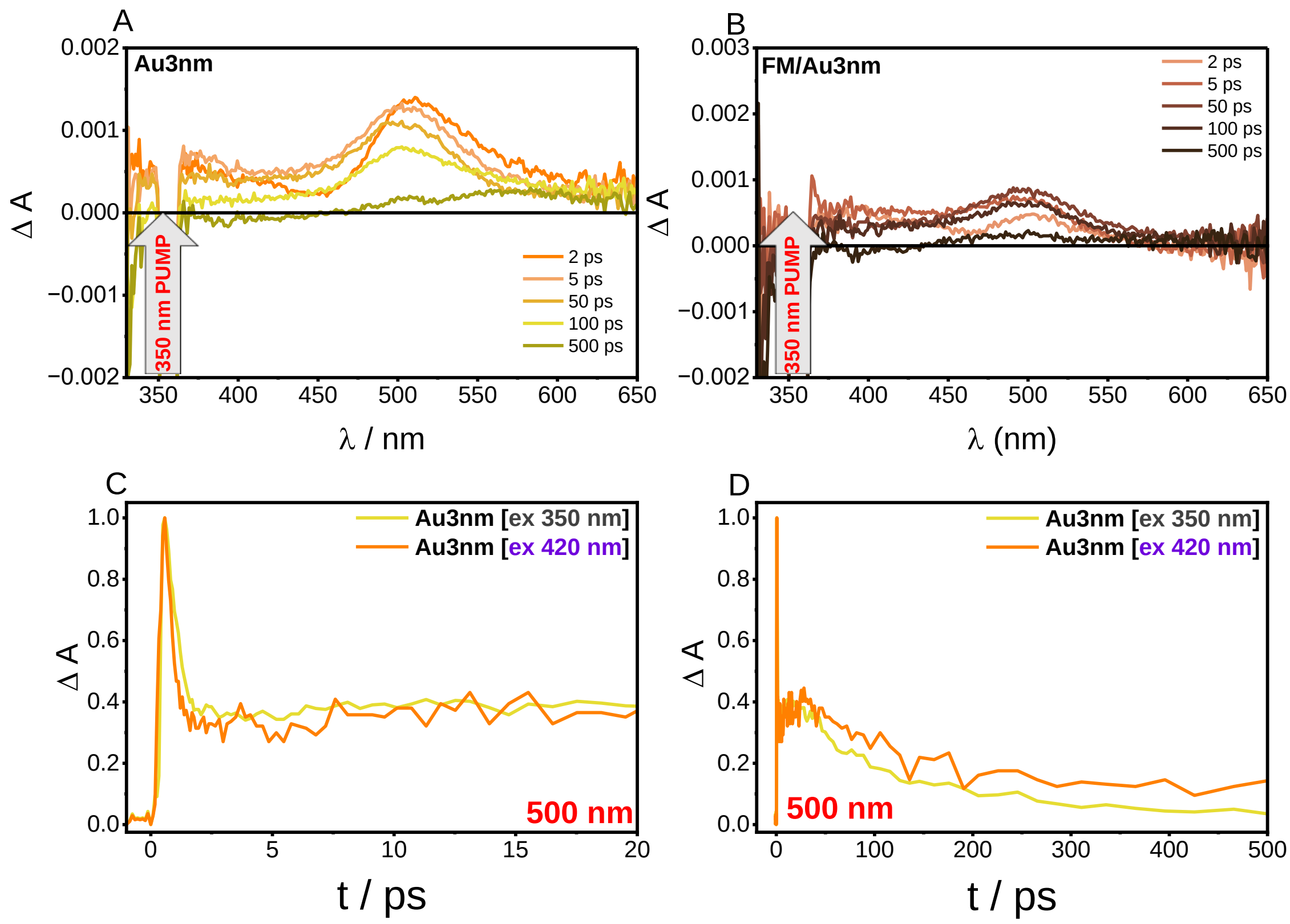

**Figure S11** Transient absorption spectra registered at various time delays for: A) 3 nm Au sample and B) FM/Au3nm sample following 350 nm excitation; C) and D) transient absorption decays at 500 nm registered for samples without FM, bearing 3 nm Au cap following excitation with either 350 nm or 420 nm pump.

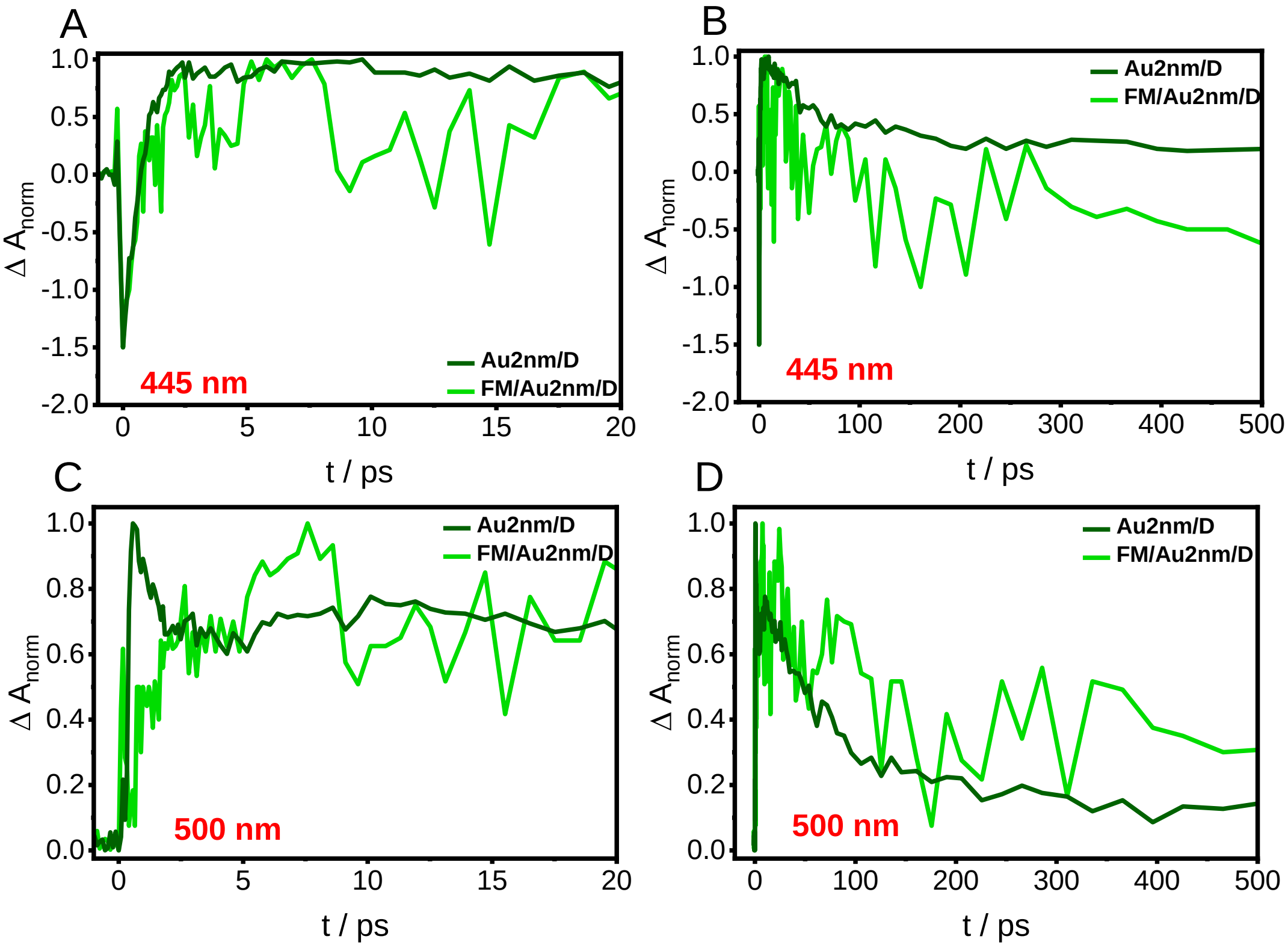

**Figure S12** Transient absorption decays at 445 nm and 500 nm on short- and long-time scale for metallic heterostructures with D-porph anchored to the gold surface following the 350 nm laser excitation. Samples with ferromagnetic layer were magnetized “up”.